\documentclass[longauth]{aa}

\usepackage{graphicx}
\usepackage{natbib}
\usepackage{scalerel}
\usepackage{amsmath}	
\usepackage{bm}	
\usepackage{amssymb}	
\usepackage{multirow}	
\usepackage{soul}

\usepackage[table]{xcolor}

\usepackage{txfonts}
\usepackage[pdfencoding=auto,psdextra]{hyperref}
\hypersetup{
    colorlinks=true,
    linkcolor=blue,
    filecolor=magenta,      
    urlcolor=blue,
    citecolor=blue
}
\usepackage{lastpage}
\makeatletter
\renewcommand*\aa@pageof{, page \thepage{} of \pageref*{LastPage}}
\makeatother

\newcommand{\ncen}{N_{\rm cen}}
\newcommand{\nsat}{N_{\rm sat}}
\newcommand{\mvir}{\ensuremath{M_{\rm vir}}\xspace}
\newcommand{\ev}[1]{\left\langle #1 \right\rangle}

\newcommand{\lcdm}{$\Lambda$CDM\xspace}
\newcommand{\wcdm}{$w$CDM\xspace}
\newcommand{\wwcdm}{$w_0w_a$CDM\xspace}

\newcommand{\fnl}[0]{f_{\rm NL}\xspace}
\newcommand{\xv}{\bm{x}}
\newcommand{\kv}{\bm{k}}

\newcommand{\mpcoh}{\, \si{\hMpc}}
\newcommand{\gpcoh}{\, h^{-1}\mathrm{Gpc}}

\newcommand{\homopc}{\, h\, \mathrm{Mpc}^{-1}}
\newcommand{\msoh}{\, h^{-1}\,M_\odot}

\newcommand{\neff}{N_{\rm eff}}

\newcommand{\Omegam}{\Omega_\mathrm{m}}
\newcommand{\vdg}{VDG$_\infty$\xspace}

\newcommand{\getdist}{\texttt{GetDist}\xspace}
\newcommand{\comet}{\texttt{COMET}\xspace}
\newcommand{\rockstar}{\texttt{ROCKSTAR}\xspace}
\newcommand{\scipic}{\texttt{SciPic}\xspace}
\newcommand{\corrfunc}{\texttt{Corrfunc}\xspace}
\newcommand{\multinest}{\texttt{MultiNest}\xspace}

\newcommand{\RAMSES}{\texttt{RAMSES}\xspace}
\newcommand{\FASTPM}{\texttt{FastPM}\xspace}

\newcommand{\LGADGETTWO}{\texttt{L-GADGET2}\xspace}
\newcommand{\GADGETTWO}{\texttt{GADGET2}\xspace}
\newcommand{\GADGETFOUR}{\texttt{GADGET4}\xspace}

\newcommand{\GIZMO}{\texttt{GIZMO}\xspace}

\newcommand{\NGENIC}{\texttt{N-GenIC}\xspace}
\newcommand{\CAMB}{\texttt{CAMB}\xspace}

\newcommand{\FLAGSHIPTWO}{\textsc{Flagship 2}\xspace}
\newcommand{\COMPLEMENTARY}{\textsc{Complementary}\xspace}
\newcommand{\DEMNUni}{\textsc{DEMNUni}\xspace}
\newcommand{\RAYGAL}{\textsc{RayGal}\xspace}

\newcommand{\DUSTGRAINPATHFINDER}{\textsc{DUSTGRAIN-PF}\xspace}
\newcommand{\CIDER}{\textsc{CiDER}\xspace}
\newcommand{\DAKAR}{\textsc{DAKAR}\xspace}
\newcommand{\DAKARTWO}{\textsc{DAKAR2}\xspace}

\newcommand{\PNGUNITsim}{\textsc{PNG-UNIT}\xspace}
\newcommand{\UNITsims}{\textsc{UNITsims}\xspace}
\newcommand{\ff}{{\mathcal F}}

\usepackage[utf8]{inputenc}

\usepackage[switch, modulo]{lineno}
\renewcommand{\linenumbers}[0]{}

\usepackage{euclid}

\begin{document}
%
%

   \title{\Euclid preparation}
   \subtitle{Simulated galaxy catalogues for non-standard cosmological models}

\newcommand{\orcid}[1]{} 
\author{Euclid Collaboration: M.-A.~Breton\thanks{\email{michel-andres.breton@obspm.fr}}\inst{\ref{aff1}}
\and P.~Fosalba\orcid{0000-0002-1510-5214}\inst{\ref{aff2},\ref{aff3}}
\and S.~Avila\orcid{0000-0001-5043-3662}\inst{\ref{aff4}}
\and M.~Baldi\orcid{0000-0003-4145-1943}\inst{\ref{aff5},\ref{aff6},\ref{aff7}}
\and C.~Carbone\orcid{0000-0003-0125-3563}\inst{\ref{aff8}}
\and M.~K{\"a}rcher\orcid{0000-0001-5868-647X}\inst{\ref{aff9}}
\and G.~R\'acz\orcid{0000-0003-3906-5699}\inst{\ref{aff10}}
\and M.~Bolzonella\orcid{0000-0003-3278-4607}\inst{\ref{aff6}}
\and F.~J.~Castander\orcid{0000-0001-7316-4573}\inst{\ref{aff3},\ref{aff2}}
\and C.~Giocoli\orcid{0000-0002-9590-7961}\inst{\ref{aff6},\ref{aff7}}
\and K.~Koyama\orcid{0000-0001-6727-6915}\inst{\ref{aff11}}
\and A.~M.~C.~Le~Brun\orcid{0000-0002-0936-4594}\inst{\ref{aff12}}
\and L.~Pozzetti\orcid{0000-0001-7085-0412}\inst{\ref{aff6}}
\and A.~G.~Adame\orcid{0009-0005-0594-9391}\inst{\ref{aff13},\ref{aff14},\ref{aff15},\ref{aff16}}
\and V.~Gonzalez-Perez\orcid{0000-0001-9938-2755}\inst{\ref{aff14},\ref{aff16}}
\and G.~Yepes\orcid{0000-0001-5031-7936}\inst{\ref{aff14},\ref{aff16}}
\and B.~Altieri\orcid{0000-0003-3936-0284}\inst{\ref{aff17}}
\and S.~Andreon\orcid{0000-0002-2041-8784}\inst{\ref{aff18}}
\and C.~Baccigalupi\orcid{0000-0002-8211-1630}\inst{\ref{aff19},\ref{aff20},\ref{aff21},\ref{aff22}}
\and S.~Bardelli\orcid{0000-0002-8900-0298}\inst{\ref{aff6}}
\and P.~Battaglia\orcid{0000-0002-7337-5909}\inst{\ref{aff6}}
\and A.~Biviano\orcid{0000-0002-0857-0732}\inst{\ref{aff20},\ref{aff19}}
\and E.~Branchini\orcid{0000-0002-0808-6908}\inst{\ref{aff23},\ref{aff24},\ref{aff18}}
\and M.~Brescia\orcid{0000-0001-9506-5680}\inst{\ref{aff25},\ref{aff26}}
\and S.~Camera\orcid{0000-0003-3399-3574}\inst{\ref{aff27},\ref{aff28},\ref{aff29}}
\and V.~Capobianco\orcid{0000-0002-3309-7692}\inst{\ref{aff29}}
\and V.~F.~Cardone\inst{\ref{aff30},\ref{aff31}}
\and J.~Carretero\orcid{0000-0002-3130-0204}\inst{\ref{aff4},\ref{aff32}}
\and M.~Castellano\orcid{0000-0001-9875-8263}\inst{\ref{aff30}}
\and G.~Castignani\orcid{0000-0001-6831-0687}\inst{\ref{aff6}}
\and S.~Cavuoti\orcid{0000-0002-3787-4196}\inst{\ref{aff26},\ref{aff33}}
\and A.~Cimatti\inst{\ref{aff34}}
\and C.~Colodro-Conde\inst{\ref{aff35}}
\and G.~Congedo\orcid{0000-0003-2508-0046}\inst{\ref{aff36}}
\and L.~Conversi\orcid{0000-0002-6710-8476}\inst{\ref{aff37},\ref{aff17}}
\and Y.~Copin\orcid{0000-0002-5317-7518}\inst{\ref{aff38}}
\and A.~Costille\inst{\ref{aff39}}
\and F.~Courbin\orcid{0000-0003-0758-6510}\inst{\ref{aff40},\ref{aff41},\ref{aff42}}
\and H.~M.~Courtois\orcid{0000-0003-0509-1776}\inst{\ref{aff43}}
\and A.~Da~Silva\orcid{0000-0002-6385-1609}\inst{\ref{aff44},\ref{aff45}}
\and H.~Degaudenzi\orcid{0000-0002-5887-6799}\inst{\ref{aff46}}
\and S.~de~la~Torre\inst{\ref{aff39}}
\and G.~De~Lucia\orcid{0000-0002-6220-9104}\inst{\ref{aff20}}
\and H.~Dole\orcid{0000-0002-9767-3839}\inst{\ref{aff47}}
\and M.~Douspis\orcid{0000-0003-4203-3954}\inst{\ref{aff47}}
\and F.~Dubath\orcid{0000-0002-6533-2810}\inst{\ref{aff46}}
\and C.~A.~J.~Duncan\orcid{0009-0003-3573-0791}\inst{\ref{aff36}}
\and X.~Dupac\inst{\ref{aff17}}
\and S.~Dusini\orcid{0000-0002-1128-0664}\inst{\ref{aff48}}
\and S.~Escoffier\orcid{0000-0002-2847-7498}\inst{\ref{aff49}}
\and M.~Farina\orcid{0000-0002-3089-7846}\inst{\ref{aff50}}
\and R.~Farinelli\inst{\ref{aff6}}
\and S.~Farrens\orcid{0000-0002-9594-9387}\inst{\ref{aff1}}
\and F.~Faustini\orcid{0000-0001-6274-5145}\inst{\ref{aff30},\ref{aff51}}
\and S.~Ferriol\inst{\ref{aff38}}
\and F.~Finelli\orcid{0000-0002-6694-3269}\inst{\ref{aff6},\ref{aff52}}
\and S.~Fotopoulou\orcid{0000-0002-9686-254X}\inst{\ref{aff53}}
\and N.~Fourmanoit\orcid{0009-0005-6816-6925}\inst{\ref{aff49}}
\and M.~Frailis\orcid{0000-0002-7400-2135}\inst{\ref{aff20}}
\and E.~Franceschi\orcid{0000-0002-0585-6591}\inst{\ref{aff6}}
\and M.~Fumana\orcid{0000-0001-6787-5950}\inst{\ref{aff8}}
\and S.~Galeotta\orcid{0000-0002-3748-5115}\inst{\ref{aff20}}
\and K.~George\orcid{0000-0002-1734-8455}\inst{\ref{aff54}}
\and B.~Gillis\orcid{0000-0002-4478-1270}\inst{\ref{aff36}}
\and J.~Gracia-Carpio\orcid{0000-0003-4689-3134}\inst{\ref{aff55}}
\and A.~Grazian\orcid{0000-0002-5688-0663}\inst{\ref{aff56}}
\and F.~Grupp\inst{\ref{aff55},\ref{aff57}}
\and S.~V.~H.~Haugan\orcid{0000-0001-9648-7260}\inst{\ref{aff58}}
\and W.~Holmes\inst{\ref{aff59}}
\and F.~Hormuth\inst{\ref{aff60}}
\and A.~Hornstrup\orcid{0000-0002-3363-0936}\inst{\ref{aff61},\ref{aff62}}
\and K.~Jahnke\orcid{0000-0003-3804-2137}\inst{\ref{aff63}}
\and M.~Jhabvala\inst{\ref{aff64}}
\and B.~Joachimi\orcid{0000-0001-7494-1303}\inst{\ref{aff65}}
\and S.~Kermiche\orcid{0000-0002-0302-5735}\inst{\ref{aff49}}
\and A.~Kiessling\orcid{0000-0002-2590-1273}\inst{\ref{aff59}}
\and M.~Kilbinger\orcid{0000-0001-9513-7138}\inst{\ref{aff1}}
\and B.~Kubik\orcid{0009-0006-5823-4880}\inst{\ref{aff38}}
\and M.~Kunz\orcid{0000-0002-3052-7394}\inst{\ref{aff66}}
\and H.~Kurki-Suonio\orcid{0000-0002-4618-3063}\inst{\ref{aff10},\ref{aff67}}
\and S.~Ligori\orcid{0000-0003-4172-4606}\inst{\ref{aff29}}
\and P.~B.~Lilje\orcid{0000-0003-4324-7794}\inst{\ref{aff58}}
\and V.~Lindholm\orcid{0000-0003-2317-5471}\inst{\ref{aff10},\ref{aff67}}
\and I.~Lloro\orcid{0000-0001-5966-1434}\inst{\ref{aff68}}
\and G.~Mainetti\orcid{0000-0003-2384-2377}\inst{\ref{aff69}}
\and O.~Mansutti\orcid{0000-0001-5758-4658}\inst{\ref{aff20}}
\and O.~Marggraf\orcid{0000-0001-7242-3852}\inst{\ref{aff70}}
\and M.~Martinelli\orcid{0000-0002-6943-7732}\inst{\ref{aff30},\ref{aff31}}
\and N.~Martinet\orcid{0000-0003-2786-7790}\inst{\ref{aff39}}
\and F.~Marulli\orcid{0000-0002-8850-0303}\inst{\ref{aff71},\ref{aff6},\ref{aff7}}
\and R.~J.~Massey\orcid{0000-0002-6085-3780}\inst{\ref{aff72}}
\and E.~Medinaceli\orcid{0000-0002-4040-7783}\inst{\ref{aff6}}
\and S.~Mei\orcid{0000-0002-2849-559X}\inst{\ref{aff73},\ref{aff74}}
\and M.~Meneghetti\orcid{0000-0003-1225-7084}\inst{\ref{aff6},\ref{aff7}}
\and E.~Merlin\orcid{0000-0001-6870-8900}\inst{\ref{aff30}}
\and G.~Meylan\inst{\ref{aff75}}
\and A.~Mora\orcid{0000-0002-1922-8529}\inst{\ref{aff76}}
\and M.~Moresco\orcid{0000-0002-7616-7136}\inst{\ref{aff71},\ref{aff6}}
\and L.~Moscardini\orcid{0000-0002-3473-6716}\inst{\ref{aff71},\ref{aff6},\ref{aff7}}
\and C.~Neissner\orcid{0000-0001-8524-4968}\inst{\ref{aff77},\ref{aff32}}
\and S.-M.~Niemi\orcid{0009-0005-0247-0086}\inst{\ref{aff78}}
\and J.~W.~Nightingale\orcid{0000-0002-8987-7401}\inst{\ref{aff79}}
\and C.~Padilla\orcid{0000-0001-7951-0166}\inst{\ref{aff77}}
\and S.~Paltani\orcid{0000-0002-8108-9179}\inst{\ref{aff46}}
\and F.~Pasian\orcid{0000-0002-4869-3227}\inst{\ref{aff20}}
\and K.~Pedersen\inst{\ref{aff80}}
\and V.~Pettorino\orcid{0000-0002-4203-9320}\inst{\ref{aff78}}
\and S.~Pires\orcid{0000-0002-0249-2104}\inst{\ref{aff1}}
\and G.~Polenta\orcid{0000-0003-4067-9196}\inst{\ref{aff51}}
\and M.~Poncet\inst{\ref{aff81}}
\and L.~A.~Popa\inst{\ref{aff82}}
\and F.~Raison\orcid{0000-0002-7819-6918}\inst{\ref{aff55}}
\and A.~Renzi\orcid{0000-0001-9856-1970}\inst{\ref{aff83},\ref{aff48},\ref{aff6}}
\and J.~Rhodes\orcid{0000-0002-4485-8549}\inst{\ref{aff59}}
\and G.~Riccio\inst{\ref{aff26}}
\and E.~Romelli\orcid{0000-0003-3069-9222}\inst{\ref{aff20}}
\and M.~Roncarelli\orcid{0000-0001-9587-7822}\inst{\ref{aff6}}
\and R.~Saglia\orcid{0000-0003-0378-7032}\inst{\ref{aff57},\ref{aff55}}
\and Z.~Sakr\orcid{0000-0002-4823-3757}\inst{\ref{aff15},\ref{aff84},\ref{aff85}}
\and D.~Sapone\orcid{0000-0001-7089-4503}\inst{\ref{aff86}}
\and B.~Sartoris\orcid{0000-0003-1337-5269}\inst{\ref{aff57},\ref{aff20}}
\and A.~Secroun\orcid{0000-0003-0505-3710}\inst{\ref{aff49}}
\and G.~Seidel\orcid{0000-0003-2907-353X}\inst{\ref{aff63}}
\and E.~Sihvola\orcid{0000-0003-1804-7715}\inst{\ref{aff87}}
\and P.~Simon\inst{\ref{aff70}}
\and C.~Sirignano\orcid{0000-0002-0995-7146}\inst{\ref{aff83},\ref{aff48}}
\and G.~Sirri\orcid{0000-0003-2626-2853}\inst{\ref{aff7}}
\and A.~Spurio~Mancini\orcid{0000-0001-5698-0990}\inst{\ref{aff88}}
\and L.~Stanco\orcid{0000-0002-9706-5104}\inst{\ref{aff48}}
\and P.~Tallada-Cresp\'{i}\orcid{0000-0002-1336-8328}\inst{\ref{aff4},\ref{aff32}}
\and A.~N.~Taylor\inst{\ref{aff36}}
\and I.~Tereno\orcid{0000-0002-4537-6218}\inst{\ref{aff44},\ref{aff89}}
\and N.~Tessore\orcid{0000-0002-9696-7931}\inst{\ref{aff90}}
\and S.~Toft\orcid{0000-0003-3631-7176}\inst{\ref{aff91},\ref{aff92}}
\and R.~Toledo-Moreo\orcid{0000-0002-2997-4859}\inst{\ref{aff93}}
\and F.~Torradeflot\orcid{0000-0003-1160-1517}\inst{\ref{aff32},\ref{aff4}}
\and I.~Tutusaus\orcid{0000-0002-3199-0399}\inst{\ref{aff3},\ref{aff2},\ref{aff84}}
\and E.~A.~Valentijn\inst{\ref{aff94}}
\and J.~Valiviita\orcid{0000-0001-6225-3693}\inst{\ref{aff10},\ref{aff67}}
\and T.~Vassallo\orcid{0000-0001-6512-6358}\inst{\ref{aff20},\ref{aff54}}
\and G.~Verdoes~Kleijn\orcid{0000-0001-5803-2580}\inst{\ref{aff94}}
\and Y.~Wang\orcid{0000-0002-4749-2984}\inst{\ref{aff95}}
\and J.~Weller\orcid{0000-0002-8282-2010}\inst{\ref{aff57},\ref{aff55}}
\and G.~Zamorani\orcid{0000-0002-2318-301X}\inst{\ref{aff6}}
\and F.~M.~Zerbi\orcid{0000-0002-9996-973X}\inst{\ref{aff18}}
\and E.~Zucca\orcid{0000-0002-5845-8132}\inst{\ref{aff6}}
\and M.~Ballardini\orcid{0000-0003-4481-3559}\inst{\ref{aff96},\ref{aff97},\ref{aff6}}
\and A.~Boucaud\orcid{0000-0001-7387-2633}\inst{\ref{aff73}}
\and E.~Bozzo\orcid{0000-0002-8201-1525}\inst{\ref{aff46}}
\and C.~Burigana\orcid{0000-0002-3005-5796}\inst{\ref{aff98},\ref{aff52}}
\and R.~Cabanac\orcid{0000-0001-6679-2600}\inst{\ref{aff84}}
\and M.~Calabrese\orcid{0000-0002-2637-2422}\inst{\ref{aff99},\ref{aff8}}
\and A.~Cappi\inst{\ref{aff100},\ref{aff6}}
\and T.~Castro\orcid{0000-0002-6292-3228}\inst{\ref{aff20},\ref{aff21},\ref{aff19},\ref{aff101}}
\and J.~A.~Escartin~Vigo\inst{\ref{aff55}}
\and G.~Fabbian\orcid{0000-0002-3255-4695}\inst{\ref{aff47}}
\and L.~Gabarra\orcid{0000-0002-8486-8856}\inst{\ref{aff102}}
\and J.~Garc\'ia-Bellido\orcid{0000-0002-9370-8360}\inst{\ref{aff15}}
\and S.~Hemmati\orcid{0000-0003-2226-5395}\inst{\ref{aff95}}
\and J.~Macias-Perez\orcid{0000-0002-5385-2763}\inst{\ref{aff103}}
\and R.~Maoli\orcid{0000-0002-6065-3025}\inst{\ref{aff104},\ref{aff30}}
\and J.~Mart\'{i}n-Fleitas\orcid{0000-0002-8594-569X}\inst{\ref{aff105}}
\and N.~Mauri\orcid{0000-0001-8196-1548}\inst{\ref{aff34},\ref{aff7}}
\and R.~B.~Metcalf\orcid{0000-0003-3167-2574}\inst{\ref{aff71},\ref{aff6}}
\and P.~Monaco\orcid{0000-0003-2083-7564}\inst{\ref{aff106},\ref{aff20},\ref{aff21},\ref{aff19}}
\and A.~Pezzotta\orcid{0000-0003-0726-2268}\inst{\ref{aff18}}
\and M.~P\"ontinen\orcid{0000-0001-5442-2530}\inst{\ref{aff10}}
\and I.~Risso\orcid{0000-0003-2525-7761}\inst{\ref{aff18},\ref{aff24}}
\and V.~Scottez\orcid{0009-0008-3864-940X}\inst{\ref{aff107},\ref{aff108}}
\and M.~Sereno\orcid{0000-0003-0302-0325}\inst{\ref{aff6},\ref{aff7}}
\and M.~Tenti\orcid{0000-0002-4254-5901}\inst{\ref{aff7}}
\and M.~Tucci\inst{\ref{aff46}}
\and M.~Viel\orcid{0000-0002-2642-5707}\inst{\ref{aff19},\ref{aff20},\ref{aff22},\ref{aff21},\ref{aff101}}
\and M.~Wiesmann\orcid{0009-0000-8199-5860}\inst{\ref{aff58}}
\and Y.~Akrami\orcid{0000-0002-2407-7956}\inst{\ref{aff15},\ref{aff109}}
\and I.~T.~Andika\orcid{0000-0001-6102-9526}\inst{\ref{aff54}}
\and G.~Angora\orcid{0000-0002-0316-6562}\inst{\ref{aff26},\ref{aff96}}
\and M.~Archidiacono\orcid{0000-0003-4952-9012}\inst{\ref{aff9},\ref{aff110}}
\and F.~Atrio-Barandela\orcid{0000-0002-2130-2513}\inst{\ref{aff111}}
\and L.~Bazzanini\orcid{0000-0003-0727-0137}\inst{\ref{aff96},\ref{aff6}}
\and J.~Bel\inst{\ref{aff112}}
\and D.~Bertacca\orcid{0000-0002-2490-7139}\inst{\ref{aff83},\ref{aff56},\ref{aff48}}
\and M.~Bethermin\orcid{0000-0002-3915-2015}\inst{\ref{aff113}}
\and F.~Beutler\orcid{0000-0003-0467-5438}\inst{\ref{aff36}}
\and A.~Blanchard\orcid{0000-0001-8555-9003}\inst{\ref{aff84}}
\and L.~Blot\orcid{0000-0002-9622-7167}\inst{\ref{aff114},\ref{aff12}}
\and M.~Bonici\orcid{0000-0002-8430-126X}\inst{\ref{aff115},\ref{aff8}}
\and S.~Borgani\orcid{0000-0001-6151-6439}\inst{\ref{aff106},\ref{aff19},\ref{aff20},\ref{aff21},\ref{aff101}}
\and M.~L.~Brown\orcid{0000-0002-0370-8077}\inst{\ref{aff116}}
\and S.~Bruton\orcid{0000-0002-6503-5218}\inst{\ref{aff117}}
\and B.~Camacho~Quevedo\orcid{0000-0002-8789-4232}\inst{\ref{aff19},\ref{aff22},\ref{aff20}}
\and F.~Caro\orcid{0009-0003-1053-0507}\inst{\ref{aff30}}
\and C.~S.~Carvalho\inst{\ref{aff89}}
\and F.~Cogato\orcid{0000-0003-4632-6113}\inst{\ref{aff71},\ref{aff6}}
\and A.~R.~Cooray\orcid{0000-0002-3892-0190}\inst{\ref{aff118}}
\and S.~Davini\orcid{0000-0003-3269-1718}\inst{\ref{aff24}}
\and G.~Desprez\orcid{0000-0001-8325-1742}\inst{\ref{aff94}}
\and A.~D\'iaz-S\'anchez\orcid{0000-0003-0748-4768}\inst{\ref{aff119}}
\and S.~Di~Domizio\orcid{0000-0003-2863-5895}\inst{\ref{aff23},\ref{aff24}}
\and J.~M.~Diego\orcid{0000-0001-9065-3926}\inst{\ref{aff120}}
\and V.~Duret\orcid{0009-0009-0383-4960}\inst{\ref{aff49}}
\and A.~Eggemeier\orcid{0000-0002-1841-8910}\inst{\ref{aff70}}
\and M.~Y.~Elkhashab\orcid{0000-0001-9306-2603}\inst{\ref{aff106},\ref{aff20},\ref{aff21},\ref{aff19}}
\and A.~Enia\orcid{0000-0002-0200-2857}\inst{\ref{aff6}}
\and Y.~Fang\orcid{0000-0002-0334-6950}\inst{\ref{aff57}}
\and A.~Finoguenov\orcid{0000-0002-4606-5403}\inst{\ref{aff10}}
\and F.~Fontanot\orcid{0000-0003-4744-0188}\inst{\ref{aff20},\ref{aff19}}
\and A.~Franco\orcid{0000-0002-4761-366X}\inst{\ref{aff121},\ref{aff122},\ref{aff123}}
\and K.~Ganga\orcid{0000-0001-8159-8208}\inst{\ref{aff73}}
\and T.~Gasparetto\orcid{0000-0002-7913-4866}\inst{\ref{aff30}}
\and R.~Gavazzi\orcid{0000-0002-5540-6935}\inst{\ref{aff39},\ref{aff124}}
\and E.~Gaztanaga\orcid{0000-0001-9632-0815}\inst{\ref{aff3},\ref{aff2},\ref{aff11}}
\and F.~Giacomini\orcid{0000-0002-3129-2814}\inst{\ref{aff7}}
\and F.~Gianotti\orcid{0000-0003-4666-119X}\inst{\ref{aff6}}
\and G.~Gozaliasl\orcid{0000-0002-0236-919X}\inst{\ref{aff125},\ref{aff10}}
\and A.~Gruppuso\orcid{0000-0001-9272-5292}\inst{\ref{aff6},\ref{aff7}}
\and M.~Guidi\orcid{0000-0001-9408-1101}\inst{\ref{aff5},\ref{aff6}}
\and C.~M.~Gutierrez\orcid{0000-0001-7854-783X}\inst{\ref{aff35},\ref{aff126}}
\and A.~Hall\orcid{0000-0002-3139-8651}\inst{\ref{aff36}}
\and H.~Hildebrandt\orcid{0000-0002-9814-3338}\inst{\ref{aff127}}
\and J.~Hjorth\orcid{0000-0002-4571-2306}\inst{\ref{aff80}}
\and J.~J.~E.~Kajava\orcid{0000-0002-3010-8333}\inst{\ref{aff128},\ref{aff129},\ref{aff130}}
\and Y.~Kang\orcid{0009-0000-8588-7250}\inst{\ref{aff46}}
\and V.~Kansal\orcid{0000-0002-4008-6078}\inst{\ref{aff131},\ref{aff132}}
\and D.~Karagiannis\orcid{0000-0002-4927-0816}\inst{\ref{aff96},\ref{aff133}}
\and K.~Kiiveri\inst{\ref{aff87}}
\and J.~Kim\orcid{0000-0003-2776-2761}\inst{\ref{aff102}}
\and C.~C.~Kirkpatrick\inst{\ref{aff87}}
\and S.~Kruk\orcid{0000-0001-8010-8879}\inst{\ref{aff17}}
\and M.~Lattanzi\orcid{0000-0003-1059-2532}\inst{\ref{aff97}}
\and J.~Le~Graet\orcid{0000-0001-6523-7971}\inst{\ref{aff49}}
\and L.~Legrand\orcid{0000-0003-0610-5252}\inst{\ref{aff134},\ref{aff135}}
\and M.~Lembo\orcid{0000-0002-5271-5070}\inst{\ref{aff124}}
\and F.~Lepori\orcid{0009-0000-5061-7138}\inst{\ref{aff136}}
\and G.~Leroy\orcid{0009-0004-2523-4425}\inst{\ref{aff137},\ref{aff72}}
\and G.~F.~Lesci\orcid{0000-0002-4607-2830}\inst{\ref{aff71},\ref{aff6}}
\and J.~Lesgourgues\orcid{0000-0001-7627-353X}\inst{\ref{aff138}}
\and T.~I.~Liaudat\orcid{0000-0002-9104-314X}\inst{\ref{aff139}}
\and S.~J.~Liu\orcid{0000-0001-7680-2139}\inst{\ref{aff50}}
\and X.~Lopez~Lopez\orcid{0009-0008-5194-5908}\inst{\ref{aff6}}
\and M.~Magliocchetti\orcid{0000-0001-9158-4838}\inst{\ref{aff50}}
\and A.~Manj\'on-Garc\'ia\orcid{0000-0002-7413-8825}\inst{\ref{aff119}}
\and F.~Mannucci\orcid{0000-0002-4803-2381}\inst{\ref{aff140}}
\and C.~J.~A.~P.~Martins\orcid{0000-0002-4886-9261}\inst{\ref{aff141},\ref{aff142}}
\and L.~Maurin\orcid{0000-0002-8406-0857}\inst{\ref{aff47}}
\and M.~Miluzio\inst{\ref{aff17},\ref{aff143}}
\and A.~Montoro\orcid{0000-0003-4730-8590}\inst{\ref{aff3},\ref{aff2}}
\and C.~Moretti\orcid{0000-0003-3314-8936}\inst{\ref{aff20},\ref{aff19},\ref{aff21}}
\and G.~Morgante\inst{\ref{aff6}}
\and S.~Nadathur\orcid{0000-0001-9070-3102}\inst{\ref{aff11}}
\and K.~Naidoo\orcid{0000-0002-9182-1802}\inst{\ref{aff11},\ref{aff63}}
\and A.~Navarro-Alsina\orcid{0000-0002-3173-2592}\inst{\ref{aff70}}
\and S.~Nesseris\orcid{0000-0002-0567-0324}\inst{\ref{aff15}}
\and L.~Pagano\orcid{0000-0003-1820-5998}\inst{\ref{aff96},\ref{aff97}}
\and D.~Paoletti\orcid{0000-0003-4761-6147}\inst{\ref{aff6},\ref{aff52}}
\and F.~Passalacqua\orcid{0000-0002-8606-4093}\inst{\ref{aff83},\ref{aff48}}
\and K.~Paterson\orcid{0000-0001-8340-3486}\inst{\ref{aff63}}
\and L.~Patrizii\inst{\ref{aff7}}
\and C.~Pattison\orcid{0000-0003-3272-2617}\inst{\ref{aff11}}
\and A.~Pisani\orcid{0000-0002-6146-4437}\inst{\ref{aff49}}
\and D.~Potter\orcid{0000-0002-0757-5195}\inst{\ref{aff144}}
\and G.~W.~Pratt\inst{\ref{aff1}}
\and S.~Quai\orcid{0000-0002-0449-8163}\inst{\ref{aff71},\ref{aff6}}
\and M.~Radovich\orcid{0000-0002-3585-866X}\inst{\ref{aff56}}
\and K.~Rojas\orcid{0000-0003-1391-6854}\inst{\ref{aff145}}
\and W.~Roster\orcid{0000-0002-9149-6528}\inst{\ref{aff55}}
\and S.~Sacquegna\orcid{0000-0002-8433-6630}\inst{\ref{aff146}}
\and M.~Sahl\'en\orcid{0000-0003-0973-4804}\inst{\ref{aff147}}
\and D.~B.~Sanders\orcid{0000-0002-1233-9998}\inst{\ref{aff148}}
\and E.~Sarpa\orcid{0000-0002-1256-655X}\inst{\ref{aff22},\ref{aff101},\ref{aff20}}
\and A.~Schneider\orcid{0000-0001-7055-8104}\inst{\ref{aff144}}
\and M.~Schultheis\inst{\ref{aff100}}
\and D.~Sciotti\orcid{0009-0008-4519-2620}\inst{\ref{aff30},\ref{aff31}}
\and E.~Sellentin\inst{\ref{aff149},\ref{aff150}}
\and L.~C.~Smith\orcid{0000-0002-3259-2771}\inst{\ref{aff151}}
\and J.~G.~Sorce\orcid{0000-0002-2307-2432}\inst{\ref{aff152},\ref{aff47}}
\and K.~Tanidis\orcid{0000-0001-9843-5130}\inst{\ref{aff153}}
\and F.~Tarsitano\orcid{0000-0002-5919-0238}\inst{\ref{aff154},\ref{aff46}}
\and G.~Testera\inst{\ref{aff24}}
\and R.~Teyssier\orcid{0000-0001-7689-0933}\inst{\ref{aff155}}
\and S.~Tosi\orcid{0000-0002-7275-9193}\inst{\ref{aff23},\ref{aff18},\ref{aff24}}
\and A.~Troja\orcid{0000-0003-0239-4595}\inst{\ref{aff83},\ref{aff48}}
\and C.~Valieri\inst{\ref{aff7}}
\and A.~Venhola\orcid{0000-0001-6071-4564}\inst{\ref{aff156}}
\and D.~Vergani\orcid{0000-0003-0898-2216}\inst{\ref{aff6}}
\and F.~Vernizzi\orcid{0000-0003-3426-2802}\inst{\ref{aff157}}
\and G.~Verza\orcid{0000-0002-1886-8348}\inst{\ref{aff158},\ref{aff159}}
\and S.~Vinciguerra\orcid{0009-0005-4018-3184}\inst{\ref{aff39}}
\and N.~A.~Walton\orcid{0000-0003-3983-8778}\inst{\ref{aff151}}
\and A.~H.~Wright\orcid{0000-0001-7363-7932}\inst{\ref{aff127}}
\and H.~W.~Yeung\orcid{0000-0002-4993-9014}\inst{\ref{aff36}}
}

\institute{Universit\'e Paris-Saclay, Universit\'e Paris Cit\'e, CEA, CNRS, AIM, 91191, Gif-sur-Yvette, France\label{aff1}
\and
Institut d'Estudis Espacials de Catalunya (IEEC),  Edifici RDIT, Campus UPC, 08860 Castelldefels, Barcelona, Spain\label{aff2}
\and
Institute of Space Sciences (ICE, CSIC), Campus UAB, Carrer de Can Magrans, s/n, 08193 Barcelona, Spain\label{aff3}
\and
Centro de Investigaciones Energ\'eticas, Medioambientales y Tecnol\'ogicas (CIEMAT), Avenida Complutense 40, 28040 Madrid, Spain\label{aff4}
\and
Dipartimento di Fisica e Astronomia, Universit\`a di Bologna, Via Gobetti 93/2, 40129 Bologna, Italy\label{aff5}
\and
INAF-Osservatorio di Astrofisica e Scienza dello Spazio di Bologna, Via Piero Gobetti 93/3, 40129 Bologna, Italy\label{aff6}
\and
INFN-Sezione di Bologna, Viale Berti Pichat 6/2, 40127 Bologna, Italy\label{aff7}
\and
INAF-IASF Milano, Via Alfonso Corti 12, 20133 Milano, Italy\label{aff8}
\and
Dipartimento di Fisica "Aldo Pontremoli", Universit\`a degli Studi di Milano, Via Celoria 16, 20133 Milano, Italy\label{aff9}
\and
Department of Physics, P.O. Box 64, University of Helsinki, 00014 Helsinki, Finland\label{aff10}
\and
Institute of Cosmology and Gravitation, University of Portsmouth, Portsmouth PO1 3FX, UK\label{aff11}
\and
Laboratoire d'etude de l'Univers et des phenomenes eXtremes, Observatoire de Paris, Universit\'e PSL, Sorbonne Universit\'e, CNRS, 92190 Meudon, France\label{aff12}
\and
Department of Astrophysics, University of Vienna, T\"urkenschanzstrasse 17, 1180 Vienna, Austria\label{aff13}
\and
Departamento de F\'isica Te\'orica, Facultad de Ciencias, Universidad Aut\'onoma de Madrid, 28049 Cantoblanco, Madrid, Spain\label{aff14}
\and
Instituto de F\'isica Te\'orica UAM-CSIC, Campus de Cantoblanco, 28049 Madrid, Spain\label{aff15}
\and
Centro de Investigaci\'{o}n Avanzada en F\'isica Fundamental (CIAFF), Facultad de Ciencias, Universidad Aut\'{o}noma de Madrid, 28049 Madrid, Spain\label{aff16}
\and
ESAC/ESA, Camino Bajo del Castillo, s/n., Urb. Villafranca del Castillo, 28692 Villanueva de la Ca\~nada, Madrid, Spain\label{aff17}
\and
INAF-Osservatorio Astronomico di Brera, Via Brera 28, 20122 Milano, Italy\label{aff18}
\and
IFPU, Institute for Fundamental Physics of the Universe, via Beirut 2, 34151 Trieste, Italy\label{aff19}
\and
INAF-Osservatorio Astronomico di Trieste, Via G. B. Tiepolo 11, 34143 Trieste, Italy\label{aff20}
\and
INFN, Sezione di Trieste, Via Valerio 2, 34127 Trieste TS, Italy\label{aff21}
\and
SISSA, International School for Advanced Studies, Via Bonomea 265, 34136 Trieste TS, Italy\label{aff22}
\and
Dipartimento di Fisica, Universit\`a di Genova, Via Dodecaneso 33, 16146, Genova, Italy\label{aff23}
\and
INFN-Sezione di Genova, Via Dodecaneso 33, 16146, Genova, Italy\label{aff24}
\and
Department of Physics "E. Pancini", University Federico II, Via Cinthia 6, 80126, Napoli, Italy\label{aff25}
\and
INAF-Osservatorio Astronomico di Capodimonte, Via Moiariello 16, 80131 Napoli, Italy\label{aff26}
\and
Dipartimento di Fisica, Universit\`a degli Studi di Torino, Via P. Giuria 1, 10125 Torino, Italy\label{aff27}
\and
INFN-Sezione di Torino, Via P. Giuria 1, 10125 Torino, Italy\label{aff28}
\and
INAF-Osservatorio Astrofisico di Torino, Via Osservatorio 20, 10025 Pino Torinese (TO), Italy\label{aff29}
\and
INAF-Osservatorio Astronomico di Roma, Via Frascati 33, 00078 Monteporzio Catone, Italy\label{aff30}
\and
INFN-Sezione di Roma, Piazzale Aldo Moro, 2 - c/o Dipartimento di Fisica, Edificio G. Marconi, 00185 Roma, Italy\label{aff31}
\and
Port d'Informaci\'{o} Cient\'{i}fica, Campus UAB, C. Albareda s/n, 08193 Bellaterra (Barcelona), Spain\label{aff32}
\and
INFN section of Naples, Via Cinthia 6, 80126, Napoli, Italy\label{aff33}
\and
Dipartimento di Fisica e Astronomia "Augusto Righi" - Alma Mater Studiorum Universit\`a di Bologna, Viale Berti Pichat 6/2, 40127 Bologna, Italy\label{aff34}
\and
Instituto de Astrof\'{\i}sica de Canarias, E-38205 La Laguna, Tenerife, Spain\label{aff35}
\and
Institute for Astronomy, University of Edinburgh, Royal Observatory, Blackford Hill, Edinburgh EH9 3HJ, UK\label{aff36}
\and
European Space Agency/ESRIN, Largo Galileo Galilei 1, 00044 Frascati, Roma, Italy\label{aff37}
\and
Universit\'e Claude Bernard Lyon 1, CNRS/IN2P3, IP2I Lyon, UMR 5822, Villeurbanne, F-69100, France\label{aff38}
\and
Aix-Marseille Universit\'e, CNRS, CNES, LAM, Marseille, France\label{aff39}
\and
Institut de Ci\`{e}ncies del Cosmos (ICCUB), Universitat de Barcelona (IEEC-UB), Mart\'{i} i Franqu\`{e}s 1, 08028 Barcelona, Spain\label{aff40}
\and
Instituci\'o Catalana de Recerca i Estudis Avan\c{c}ats (ICREA), Passeig de Llu\'{\i}s Companys 23, 08010 Barcelona, Spain\label{aff41}
\and
Institut de Ciencies de l'Espai (IEEC-CSIC), Campus UAB, Carrer de Can Magrans, s/n Cerdanyola del Vall\'es, 08193 Barcelona, Spain\label{aff42}
\and
UCB Lyon 1, CNRS/IN2P3, IUF, IP2I Lyon, 4 rue Enrico Fermi, 69622 Villeurbanne, France\label{aff43}
\and
Departamento de F\'isica, Faculdade de Ci\^encias, Universidade de Lisboa, Edif\'icio C8, Campo Grande, PT1749-016 Lisboa, Portugal\label{aff44}
\and
Instituto de Astrof\'isica e Ci\^encias do Espa\c{c}o, Faculdade de Ci\^encias, Universidade de Lisboa, Campo Grande, 1749-016 Lisboa, Portugal\label{aff45}
\and
Department of Astronomy, University of Geneva, ch. d'Ecogia 16, 1290 Versoix, Switzerland\label{aff46}
\and
Universit\'e Paris-Saclay, CNRS, Institut d'astrophysique spatiale, 91405, Orsay, France\label{aff47}
\and
INFN-Padova, Via Marzolo 8, 35131 Padova, Italy\label{aff48}
\and
Aix-Marseille Universit\'e, CNRS/IN2P3, CPPM, Marseille, France\label{aff49}
\and
INAF-Istituto di Astrofisica e Planetologia Spaziali, via del Fosso del Cavaliere, 100, 00100 Roma, Italy\label{aff50}
\and
Space Science Data Center, Italian Space Agency, via del Politecnico snc, 00133 Roma, Italy\label{aff51}
\and
INFN-Bologna, Via Irnerio 46, 40126 Bologna, Italy\label{aff52}
\and
School of Physics, HH Wills Physics Laboratory, University of Bristol, Tyndall Avenue, Bristol, BS8 1TL, UK\label{aff53}
\and
University Observatory, LMU Faculty of Physics, Scheinerstr.~1, 81679 Munich, Germany\label{aff54}
\and
Max Planck Institute for Extraterrestrial Physics, Giessenbachstr. 1, 85748 Garching, Germany\label{aff55}
\and
INAF-Osservatorio Astronomico di Padova, Via dell'Osservatorio 5, 35122 Padova, Italy\label{aff56}
\and
Universit\"ats-Sternwarte M\"unchen, Fakult\"at f\"ur Physik, Ludwig-Maximilians-Universit\"at M\"unchen, Scheinerstr.~1, 81679 M\"unchen, Germany\label{aff57}
\and
Institute of Theoretical Astrophysics, University of Oslo, P.O. Box 1029 Blindern, 0315 Oslo, Norway\label{aff58}
\and
Jet Propulsion Laboratory, California Institute of Technology, 4800 Oak Grove Drive, Pasadena, CA, 91109, USA\label{aff59}
\and
Felix Hormuth Engineering, Goethestr. 17, 69181 Leimen, Germany\label{aff60}
\and
Technical University of Denmark, Elektrovej 327, 2800 Kgs. Lyngby, Denmark\label{aff61}
\and
Cosmic Dawn Center (DAWN), Denmark\label{aff62}
\and
Max-Planck-Institut f\"ur Astronomie, K\"onigstuhl 17, 69117 Heidelberg, Germany\label{aff63}
\and
NASA Goddard Space Flight Center, Greenbelt, MD 20771, USA\label{aff64}
\and
Department of Physics and Astronomy, University College London, Gower Street, London WC1E 6BT, UK\label{aff65}
\and
Universit\'e de Gen\`eve, D\'epartement de Physique Th\'eorique and Centre for Astroparticle Physics, 24 quai Ernest-Ansermet, CH-1211 Gen\`eve 4, Switzerland\label{aff66}
\and
Helsinki Institute of Physics, Gustaf H{\"a}llstr{\"o}min katu 2, University of Helsinki, 00014 Helsinki, Finland\label{aff67}
\and
SKAO, Jodrell Bank, Lower Withington, Macclesfield SK11 9FT, UK\label{aff68}
\and
Centre de Calcul de l'IN2P3/CNRS, 21 avenue Pierre de Coubertin 69627 Villeurbanne Cedex, France\label{aff69}
\and
Universit\"at Bonn, Argelander-Institut f\"ur Astronomie, Auf dem H\"ugel 71, 53121 Bonn, Germany\label{aff70}
\and
Dipartimento di Fisica e Astronomia "Augusto Righi" - Alma Mater Studiorum Universit\`a di Bologna, via Piero Gobetti 93/2, 40129 Bologna, Italy\label{aff71}
\and
Department of Physics, Institute for Computational Cosmology, Durham University, South Road, Durham, DH1 3LE, UK\label{aff72}
\and
Universit\'e Paris Cit\'e, CNRS, Astroparticule et Cosmologie, 75013 Paris, France\label{aff73}
\and
CNRS-UCB International Research Laboratory, Centre Pierre Bin\'etruy, IRL2007, CPB-IN2P3, Berkeley, USA\label{aff74}
\and
Institute of Physics, Laboratory of Astrophysics, Ecole Polytechnique F\'ed\'erale de Lausanne (EPFL), Observatoire de Sauverny, 1290 Versoix, Switzerland\label{aff75}
\and
Telespazio UK S.L. for European Space Agency (ESA), Camino bajo del Castillo, s/n, Urbanizacion Villafranca del Castillo, Villanueva de la Ca\~nada, 28692 Madrid, Spain\label{aff76}
\and
Institut de F\'{i}sica d'Altes Energies (IFAE), The Barcelona Institute of Science and Technology, Campus UAB, 08193 Bellaterra (Barcelona), Spain\label{aff77}
\and
European Space Agency/ESTEC, Keplerlaan 1, 2201 AZ Noordwijk, The Netherlands\label{aff78}
\and
School of Mathematics, Statistics and Physics, Newcastle University, Herschel Building, Newcastle-upon-Tyne, NE1 7RU, UK\label{aff79}
\and
DARK, Niels Bohr Institute, University of Copenhagen, Jagtvej 155, 2200 Copenhagen, Denmark\label{aff80}
\and
Centre National d'Etudes Spatiales -- Centre spatial de Toulouse, 18 avenue Edouard Belin, 31401 Toulouse Cedex 9, France\label{aff81}
\and
Institute of Space Science, Str. Atomistilor, nr. 409 M\u{a}gurele, Ilfov, 077125, Romania\label{aff82}
\and
Dipartimento di Fisica e Astronomia "G. Galilei", Universit\`a di Padova, Via Marzolo 8, 35131 Padova, Italy\label{aff83}
\and
Institut de Recherche en Astrophysique et Plan\'etologie (IRAP), Universit\'e de Toulouse, CNRS, UPS, CNES, 14 Av. Edouard Belin, 31400 Toulouse, France\label{aff84}
\and
Universit\'e St Joseph; Faculty of Sciences, Beirut, Lebanon\label{aff85}
\and
Departamento de F\'isica, FCFM, Universidad de Chile, Blanco Encalada 2008, Santiago, Chile\label{aff86}
\and
Department of Physics and Helsinki Institute of Physics, Gustaf H\"allstr\"omin katu 2, University of Helsinki, 00014 Helsinki, Finland\label{aff87}
\and
Department of Physics, Royal Holloway, University of London, Surrey TW20 0EX, UK\label{aff88}
\and
Instituto de Astrof\'isica e Ci\^encias do Espa\c{c}o, Faculdade de Ci\^encias, Universidade de Lisboa, Tapada da Ajuda, 1349-018 Lisboa, Portugal\label{aff89}
\and
Mullard Space Science Laboratory, University College London, Holmbury St Mary, Dorking, Surrey RH5 6NT, UK\label{aff90}
\and
Cosmic Dawn Center (DAWN)\label{aff91}
\and
Niels Bohr Institute, University of Copenhagen, Jagtvej 128, 2200 Copenhagen, Denmark\label{aff92}
\and
Universidad Polit\'ecnica de Cartagena, Departamento de Electr\'onica y Tecnolog\'ia de Computadoras,  Plaza del Hospital 1, 30202 Cartagena, Spain\label{aff93}
\and
Kapteyn Astronomical Institute, University of Groningen, PO Box 800, 9700 AV Groningen, The Netherlands\label{aff94}
\and
Caltech/IPAC, 1200 E. California Blvd., Pasadena, CA 91125, USA\label{aff95}
\and
Dipartimento di Fisica e Scienze della Terra, Universit\`a degli Studi di Ferrara, Via Giuseppe Saragat 1, 44122 Ferrara, Italy\label{aff96}
\and
Istituto Nazionale di Fisica Nucleare, Sezione di Ferrara, Via Giuseppe Saragat 1, 44122 Ferrara, Italy\label{aff97}
\and
INAF, Istituto di Radioastronomia, Via Piero Gobetti 101, 40129 Bologna, Italy\label{aff98}
\and
Astronomical Observatory of the Autonomous Region of the Aosta Valley (OAVdA), Loc. Lignan 39, I-11020, Nus (Aosta Valley), Italy\label{aff99}
\and
Universit\'e C\^{o}te d'Azur, Observatoire de la C\^{o}te d'Azur, CNRS, Laboratoire Lagrange, Bd de l'Observatoire, CS 34229, 06304 Nice cedex 4, France\label{aff100}
\and
ICSC - Centro Nazionale di Ricerca in High Performance Computing, Big Data e Quantum Computing, Via Magnanelli 2, Bologna, Italy\label{aff101}
\and
Department of Physics, Oxford University, Keble Road, Oxford OX1 3RH, UK\label{aff102}
\and
Univ. Grenoble Alpes, CNRS, Grenoble INP, LPSC-IN2P3, 53, Avenue des Martyrs, 38000, Grenoble, France\label{aff103}
\and
Dipartimento di Fisica, Sapienza Universit\`a di Roma, Piazzale Aldo Moro 2, 00185 Roma, Italy\label{aff104}
\and
Aurora Technology for European Space Agency (ESA), Camino bajo del Castillo, s/n, Urbanizacion Villafranca del Castillo, Villanueva de la Ca\~nada, 28692 Madrid, Spain\label{aff105}
\and
Dipartimento di Fisica - Sezione di Astronomia, Universit\`a di Trieste, Via Tiepolo 11, 34131 Trieste, Italy\label{aff106}
\and
Institut d'Astrophysique de Paris, 98bis Boulevard Arago, 75014, Paris, France\label{aff107}
\and
ICL, Junia, Universit\'e Catholique de Lille, LITL, 59000 Lille, France\label{aff108}
\and
CERCA/ISO, Department of Physics, Case Western Reserve University, 10900 Euclid Avenue, Cleveland, OH 44106, USA\label{aff109}
\and
INFN-Sezione di Milano, Via Celoria 16, 20133 Milano, Italy\label{aff110}
\and
Departamento de F{\'\i}sica Fundamental. Universidad de Salamanca. Plaza de la Merced s/n. 37008 Salamanca, Spain\label{aff111}
\and
Aix-Marseille Universit\'e, Universit\'e de Toulon, CNRS, CPT, Marseille, France\label{aff112}
\and
Universit\'e de Strasbourg, CNRS, Observatoire astronomique de Strasbourg, UMR 7550, 67000 Strasbourg, France\label{aff113}
\and
Center for Data-Driven Discovery, Kavli IPMU (WPI), UTIAS, The University of Tokyo, Kashiwa, Chiba 277-8583, Japan\label{aff114}
\and
Waterloo Centre for Astrophysics, University of Waterloo, Waterloo, Ontario N2L 3G1, Canada\label{aff115}
\and
Jodrell Bank Centre for Astrophysics, Department of Physics and Astronomy, University of Manchester, Oxford Road, Manchester M13 9PL, UK\label{aff116}
\and
California Institute of Technology, 1200 E California Blvd, Pasadena, CA 91125, USA\label{aff117}
\and
Department of Physics \& Astronomy, University of California Irvine, Irvine CA 92697, USA\label{aff118}
\and
Departamento F\'isica Aplicada, Universidad Polit\'ecnica de Cartagena, Campus Muralla del Mar, 30202 Cartagena, Murcia, Spain\label{aff119}
\and
Instituto de F\'isica de Cantabria, Edificio Juan Jord\'a, Avenida de los Castros, 39005 Santander, Spain\label{aff120}
\and
INFN, Sezione di Lecce, Via per Arnesano, CP-193, 73100, Lecce, Italy\label{aff121}
\and
Department of Mathematics and Physics E. De Giorgi, University of Salento, Via per Arnesano, CP-I93, 73100, Lecce, Italy\label{aff122}
\and
INAF-Sezione di Lecce, c/o Dipartimento Matematica e Fisica, Via per Arnesano, 73100, Lecce, Italy\label{aff123}
\and
Institut d'Astrophysique de Paris, UMR 7095, CNRS, and Sorbonne Universit\'e, 98 bis boulevard Arago, 75014 Paris, France\label{aff124}
\and
Department of Computer Science, Aalto University, PO Box 15400, Espoo, FI-00 076, Finland\label{aff125}
\and
Universidad de La Laguna, Dpto. Astrof\'\i sica, E-38206 La Laguna, Tenerife, Spain\label{aff126}
\and
Ruhr University Bochum, Faculty of Physics and Astronomy, Astronomical Institute (AIRUB), German Centre for Cosmological Lensing (GCCL), 44780 Bochum, Germany\label{aff127}
\and
Department of Physics and Astronomy, Vesilinnantie 5, University of Turku, 20014 Turku, Finland\label{aff128}
\and
Finnish Centre for Astronomy with ESO (FINCA), Quantum, Vesilinnantie 5, University of Turku, 20014 Turku, Finland\label{aff129}
\and
Serco for European Space Agency (ESA), Camino bajo del Castillo, s/n, Urbanizacion Villafranca del Castillo, Villanueva de la Ca\~nada, 28692 Madrid, Spain\label{aff130}
\and
ARC Centre of Excellence for Dark Matter Particle Physics, Melbourne, Australia\label{aff131}
\and
Centre for Astrophysics \& Supercomputing, Swinburne University of Technology,  Hawthorn, Victoria 3122, Australia\label{aff132}
\and
Department of Physics and Astronomy, University of the Western Cape, Bellville, Cape Town, 7535, South Africa\label{aff133}
\and
DAMTP, Centre for Mathematical Sciences, Wilberforce Road, Cambridge CB3 0WA, UK\label{aff134}
\and
Kavli Institute for Cosmology Cambridge, Madingley Road, Cambridge, CB3 0HA, UK\label{aff135}
\and
Departement of Theoretical Physics, University of Geneva, Switzerland\label{aff136}
\and
Department of Physics, Centre for Extragalactic Astronomy, Durham University, South Road, Durham, DH1 3LE, UK\label{aff137}
\and
Institute for Theoretical Particle Physics and Cosmology (TTK), RWTH Aachen University, 52056 Aachen, Germany\label{aff138}
\and
IRFU, CEA, Universit\'e Paris-Saclay 91191 Gif-sur-Yvette Cedex, France\label{aff139}
\and
INAF-Osservatorio Astrofisico di Arcetri, Largo E. Fermi 5, 50125, Firenze, Italy\label{aff140}
\and
Centro de Astrof\'{\i}sica da Universidade do Porto, Rua das Estrelas, 4150-762 Porto, Portugal\label{aff141}
\and
Instituto de Astrof\'isica e Ci\^encias do Espa\c{c}o, Universidade do Porto, CAUP, Rua das Estrelas, PT4150-762 Porto, Portugal\label{aff142}
\and
HE Space for European Space Agency (ESA), Camino bajo del Castillo, s/n, Urbanizacion Villafranca del Castillo, Villanueva de la Ca\~nada, 28692 Madrid, Spain\label{aff143}
\and
Department of Astrophysics, University of Zurich, Winterthurerstrasse 190, 8057 Zurich, Switzerland\label{aff144}
\and
University of Applied Sciences and Arts of Northwestern Switzerland, School of Computer Science, 5210 Windisch, Switzerland\label{aff145}
\and
INAF - Osservatorio Astronomico d'Abruzzo, Via Maggini, 64100, Teramo, Italy\label{aff146}
\and
Theoretical astrophysics, Department of Physics and Astronomy, Uppsala University, Box 516, 751 37 Uppsala, Sweden\label{aff147}
\and
Institute for Astronomy, University of Hawaii, 2680 Woodlawn Drive, Honolulu, HI 96822, USA\label{aff148}
\and
Mathematical Institute, University of Leiden, Einsteinweg 55, 2333 CA Leiden, The Netherlands\label{aff149}
\and
Leiden Observatory, Leiden University, Einsteinweg 55, 2333 CC Leiden, The Netherlands\label{aff150}
\and
Institute of Astronomy, University of Cambridge, Madingley Road, Cambridge CB3 0HA, UK\label{aff151}
\and
Univ. Lille, CNRS, Centrale Lille, UMR 9189 CRIStAL, 59000 Lille, France\label{aff152}
\and
Center for Astrophysics and Cosmology, University of Nova Gorica, Nova Gorica, Slovenia\label{aff153}
\and
Institute for Particle Physics and Astrophysics, Dept. of Physics, ETH Zurich, Wolfgang-Pauli-Strasse 27, 8093 Zurich, Switzerland\label{aff154}
\and
Department of Astrophysical Sciences, Peyton Hall, Princeton University, Princeton, NJ 08544, USA\label{aff155}
\and
Space physics and astronomy research unit, University of Oulu, Pentti Kaiteran katu 1, FI-90014 Oulu, Finland\label{aff156}
\and
Institut de Physique Th\'eorique, CEA, CNRS, Universit\'e Paris-Saclay 91191 Gif-sur-Yvette Cedex, France\label{aff157}
\and
International Centre for Theoretical Physics (ICTP), Strada Costiera 11, 34151 Trieste, Italy\label{aff158}
\and
Center for Computational Astrophysics, Flatiron Institute, 162 5th Avenue, 10010, New York, NY, USA\label{aff159}}    
%
%
%
%

%
%
   \abstract
   {Stage-IV galaxy surveys will provide the opportunity to test cosmological models and the underlying theory of gravity with unparalleled precision. 
   In this context, it is crucial for the \Euclid mission to leverage its spectroscopic and photometric probes to systematically investigate and incorporate non-standard cosmological models, including modified gravity, alternative dark energy scenarios, massive neutrinos, and primordial non-Gaussianity.
For this, we produce and release publicly simulated galaxy catalogues from a broad suite of non-standard cosmological simulations, which we processed through a model-independent analytical pipeline, making use of \rockstar for halo identification, and a modified version of the \scipic\ library for the galaxy-halo connection using the halo occupation distribution framework. We investigate their galaxy-clustering characteristics via the multipoles of the 2-point correlation function in redshift space and \vdg, a highly performant, state-of-the-art model for galaxy clustering.
Across a wide range of models, the linear growth rate multiplied by the matter density within spheres of radius 12\,Mpc, $f\sigma_{12}$, exhibits a notable robustness to the choice of cosmological template. Compared to previous works, our study extends this result to numerous scenarios with markedly distinct gravitational or dark energy dynamics. 
We find that the most of the scatter in cosmological parameter inference already appears when using the cosmological model of the simulations as templates. Using a `wrong' template can also introduce an additional scatter, although with smaller amplitude. Often, we find deviations much larger than error bars, meaning that the Gaussian approximation for the covariance might need to be further studied.
Looking ahead, future cosmological investigations must broaden their scope to include a diverse array of non-standard theoretical frameworks, extending beyond \lcdm and rudimentary dynamic dark energy models.
}
%
%
%
%
   \titlerunning{Simulated galaxy catalogues beyond \lcdm}
   \authorrunning{Euclid Collaboration: M.-A. Breton et al.}

   \maketitle
%
%
%
%
   
\section{\label{sec:Intro}Introduction}

The discovery of the accelerated \citep{perlmutter1998discovery,riess1998observational} expansion established the \lcdm model (with $\Lambda$ a cosmological constant and non-relativistic cold dark matter) as the concordance model in cosmology. Since then, this model has been confirmed by a vast majority of observational probes, with some discrepancies being characterised as statistical `tensions', such as the `$H_0$ tension' \citep[e.g.][]{verde2024tale}, where early-time measurements (e.g., the anisotropies of the cosmic microwave background, \citealt{planck2018cosmological}) yield a different value of the Hubble constant $H_0$ than late-time observations (e.g., supernovae and Cepheids, \citealt{riess2022comprehensive}). However, while potentially indicative of new physics, these discrepancies have not yet precipitated a paradigm shift.

We are currently in the era of Stage-IV surveys, including the Dark Energy Spectroscopic Instrument \citep{desi2016desi}, the Vera C. Rubin Observatory Legacy Survey of Space and Time \citep[LSST,][]{ivezic2019lsst}, the {\it Roman} Space Telescope \citep{spergel2015wfirst}, and \Euclid\ \citep{Laureijs11, EuclidSkyOverview}. These surveys are expected to significantly enhance our understanding of the Universe’s nature and assist in resolving the current tensions.
DESI, a ground-based spectroscopic experiment that began data collection in 2019, is the first operational Stage-IV survey. Its first data release \citep{desi2025constraintsBAO_DR1} demonstrated that an extended parameter-space analysis involving two parameters to describe dynamical dark energy did not support the \lcdm model beyond the 2.5\,$\sigma$ level. This result was further confirmed by its second data release \citep{desi2025constraintsBAO_DR2}. The fact that early-stage results from the first Stage-IV experiment are already capable of challenging the concordance model opens up promising and exciting avenues for future investigation.
\Euclid\ is the second of the Stage-IV surveys, launched on July 1, 2023. It will observe billions of galaxies through photometry, and of the order of 50 million through spectroscopy. For the spectroscopic part, the Euclid Wide Survey will detect galaxies with H$\alpha$ line flux above $2\times10^{-16}$\,erg\,s$^{-1}$\,cm$^{-2}$ with signal-to-noise ratio S/N\,$\geq$\,3.5, and an $H$-band magnitude $H < 24$ \citep{Laureijs11,EuclidSkyOverview}.
Its first data release is expected in 2026 and will play a critical role in either corroborating or refuting DESI's findings.

The increased precision of cosmological data and current challenges to the \lcdm model necessitate broadening the range of models we consider as part of our standard analysis pipeline. Such models must include modifications to the dark energy equation of state (beyond a cosmological constant), modifications to the theory of gravity (beyond General Relativity), as well as massive neutrinos and primordial non-Gaussianity (PNG). Indeed, constraining these deviations is the main goal of \Euclid.

A convenient probe for spectroscopic galaxy surveys is the clustering of galaxies, which is considered model-independent with the template-fitting method. It consists of computing the multipoles of the 2-point correlation function (2PCF) or power spectrum, and fitting the growth rate and distance dilations (or Alcock--Paczynski parameters, hereafter AP, \citealt{alcock1979evolution}) with a template based on fiducial cosmological parameters (because redshifts are converted into distances, assuming a fiducial cosmology). The measurement of the growth rate, in particular, has been a standard test of General Relativity (GR) at cosmological scales \citep{delatorre2013vipers, zarrouk2018clustering,bautista2021completed}.

Because we need to be prepared to analyse a Universe that is not simply described by \lcdm, we need robust analysis tools that include non-standard models. This is precisely the role of simulations, which allow us to produce realisations of universes under various theoretical assumptions in order to test our pipelines. Although some non-standard models necessitate specific numerical methods to solve nonlinear differential equations for additional fields, \cite{EP-Adamek} showed that all the currently available state-of-the-art non-standard simulation codes yield very similar results, close to the percent level agreement on the matter power spectrum, and close to 5\% on the halo mass function. Within the Euclid\ Consortium, a coordinated effort has been made to analyse a wide range of non-standard simulations and produce dark matter halo catalogues with \rockstar\ \citep{behroozi2013rockstar}, all processed through the same model-agnostic pipeline in \cite{EP-Racz}. These catalogues and other statistical outputs are available in CosmoHub \citep{carretero2017cosmohub,tallada2020cosmohub}.\footnote{\href{https://cosmohub.pic.es/}{https://cosmohub.pic.es/}, see also Sect.\,\ref{sec:data_availability}.}

In this paper, we aim to produce galaxy catalogues for a wide range of simulations and non-standard models, using the halo catalogues previously mentioned. The galaxy-halo connection is modeled using the halo-occupancy distribution (HOD) method, such that the clustering of galaxies matches that of the expected spectroscopic sample of \Euclid.
We then perform a galaxy-clustering analysis using the multipoles of the 2PCF in redshift space while maintaining a model-agnostic pipeline.

The paper is organised as follows: in Sect.\,\ref{sec:Theory} we review the different non-standard models we consider, as well as the galaxy clustering formalism. In Sect.\,\ref{sec:simulations}, we present the cosmological simulation suites we considered for the present work, and Sect.\,\ref{sec:Methods} describes the methodology to produce galaxy catalogues and their statistical analysis. We present the results in Sect.\,\ref{sec:Results} and conclude in Sect.\,\ref{sec:Conclusion}.

\section{\label{sec:Theory}Theory}

In this section, we describe the cosmological models underlying our non-standard simulations and the galaxy clustering models that we use to analyse the redshift-space 2PCF multipoles of the galaxy catalogues we produce.

\subsection{Non-standard cosmological models}

We consider `non-standard' the cosmological models that depart from vanilla \lcdm. Such models include modified gravity theories, dark energy, primordial non-Gaussianity, and massive neutrinos. The models we use have been described in \cite{EP-Racz}. Here, we aim to provide a brief description of the main features of these models.

\label{sec:non-standard_models}
\subsubsection{Massive neutrinos}
Neutrinos are mainly characterised by two properties: their total mass, $M_\nu \equiv \sum_{i=1}^{N_\nu}\,m_{\nu,i}$, with $N_\nu$ the number of neutrinos, and $\neff$ their effective number. In the standard model, $N_{\rm eff}\sim$ 3.045 \citep{Cielo:2023bqp} for three families of active neutrinos that thermalised in the early Universe and decoupled well before electron-positron annihilation. A deviation from the fiducial value accounts for the presence of non-standard neutrino features or additional relativistic relics contributing to the energy budget of the Universe \citep{Mangano:2001iu}. Here, we focus on standard neutrino families.

Oscillation experiments \citep{Maltoni:2004ei,Kajita:2016cak} showed that at least two neutrinos are massive by measuring two squared-mass differences. It can be shown that the minimum value of $M_\nu$ is either $0.06\,{\rm eV}$ in the so-called normal hierarchy or $0.10\,{\rm eV}$ in the inverted hierarchy. This value can be constrained through cosmological observations since neutrinos modify the expansion history of the Universe and suppress the clustering of dark matter, which can be observed in the large-scale distribution of galaxies \citep{Bond:1980ha,Hu:1997mj,Castorina:2015bma,Sakr:2022ans}. More specifically, we assume that the energy content of the Universe is composed of cold dark matter (CDM, with total density $\rho_{\rm c}$ and density parameter $\Omega_{\rm c}$), baryons ($\rho_{\rm b}$, $\Omega_{\rm b}$), photons ($\rho_\gamma$, $\Omega_\gamma$), neutrinos ($\rho_\nu$, $\Omega_\nu$), and a dark energy component ($\rho_{\rm DE}$, $\Omega_{\rm DE}$). We refer to the sum of CDM and baryon densities as the total `cold matter' component, with the density parameter given by $\Omega_{\rm cb}\equiv\Omega_{\rm c}+\Omega_{\rm b}$, while we distinguish a massless (relativistic) neutrino component ($\rho_\nu^{\rm r}$, $\Omega_\nu^{\rm r}$) from a massive (non-relativistic) component ($\rho_\nu^{\rm nr}$, $\Omega_\nu^{\rm nr}$), such that $\Omega_{\nu}\equiv\Omega_\nu^{\rm r}+\Omega_\nu^{\rm nr}$.
The total matter content is given by CDM, baryons, and massive neutrinos, with a density parameter $\Omega_{\rm m}\equiv\Omega_{\rm cb}+\Omega_{\nu}^{\rm nr}$. Throughout this paper, the subscript `,0'
refers to present-day quantities.

The evolution of both photon and neutrino densities depends on their momentum distributions. For photons, it can be expressed as 
\begin{equation}
\rho_\gamma(z)=\frac{\pi^2}{15}\,\frac{(k_{\rm B} T_{\gamma,0})^4}{(\hbar c)^3}\,(1+z)^4\,,
\label{densityphoton}
\end{equation}
where $k_{\rm B}$ is Boltzmann's constant, while for a neutrino species of mass $m_{\nu,i}$ in units of eV, it reads
\begin{equation}
\rho_{\nu,i}(z)=\frac{(k_{\rm B} T_{\nu,0})^4}{\pi^2 (\hbar c)^3}\,(1+z)^4\ff\left[ \frac{c^2 m_{\nu,i}}{k_B T_{\nu,0}(1+z)}\right]\,,
\label{densityneutrino}
\end{equation}
where $T_{\gamma,0}$ and $T_{\nu,0}$ are, respectively, the photon and neutrino temperatures today, and the function $\ff$ is defined as 
\begin{equation}
\ff(y)\equiv\int_0^\infty\!\!\frac{x^2\sqrt{x^2 + y^2}}{1+{\rm e}^x}\diff x\,.
\label{ff}
\end{equation}
It is convenient to express the neutrino energy density in terms of the photon density as 
\begin{equation}
\rho_{\nu,i}(z)=\frac{15}{\pi^4}\,\Gamma_{\nu}^4\,\rho_\gamma(z)\,\ff\left[ \frac{c^2 m_{\nu,i}}{k_{\rm B} T_{\nu,0}(1+z)}\right]\,,
\label{neutrino}
\end{equation}
where $\Gamma_\nu\equiv T_{\nu,0}/T_{\gamma,0}$ is the neutrino-to-photon temperature ratio today. In the limit of instantaneous decoupling, $\Gamma_{\nu,\rm inst}=\left( 4/11\right)^{1/3}$.
Under these assumptions, the evolution of the neutrino contribution to the expansion rate of the Universe is 
\begin{equation}
\Omega_\nu(z)\,E^2(z) = \frac{15}{\pi^4}\,\Gamma_\nu^4\, N_\nu\,\Omega_{\gamma,0} \,(1+z)^4\, \ff\left[ \frac{ c^2 M_\nu/ (\Gamma_\nu\, N_\nu\, k_B\, T_{\gamma,0}) }{1+z} \right],
\label{onue2}
\end{equation}
where $E(z)$ describes the time dependence of the Hubble rate, such that $H(z)\equiv H_0 E(z)$. The Hubble parameter is, therefore, given by
\begin{equation}
H(z) = H_0 \sqrt{ \Omega_{\gamma,0} (1+z)^4 + \Omega_{\rm cb,0}(1+z)^3 \\
    + \Omega_\nu (z) E^2(z) + \Omega_{\rm DE,0} }\,,
\label{hofz}
\end{equation}
where $\Omega_{\rm \gamma,0}$ represents the residual contribution of photons, given by
$\Omega_{\rm \gamma,0}\,h^2=2.469\times 10^{-5}$,
obtained from Eq.~\eqref{densityphoton} in terms of the CMB temperature, assuming $T_{\gamma,0}=2.7255$ K.\footnote{We remark that in a $\Lambda$CDM cosmology with massless neutrinos, the computation of the Hubble function, as expressed in Eq.~\eqref{hofz}, does not include the neutrino energy density parameter given by Eq.~\eqref{onue2}, but rather its relativistic limit,
$\Omega_\nu(z)E^2(z) = 7/8\,N_{\rm eff}  \left( 4/11 \right)^{4/3} \Omega_{\gamma,0} (1+z)^4$,
and therefore contributes to the radiation energy density.}

In the non-relativistic, late-time limit  $c^2 m_{\nu,i}\gg k_{\rm B}T_{\nu,0}(1+z)$, or for $z\ll z_{\rm nr}$ with the redshift of non-relativistic transition $z_{\rm nr}$ estimated as $1+z_{\rm nr} \simeq 1890 \, c^2 m_{\nu,i}/1 ~{\rm eV}$,
one obtains $\ff \rightarrow 3/2 \, y \,\zeta(3)$, where $\zeta$ is the Riemann zeta function, so that
\begin{equation}
\rho_{\nu}(z)=\frac{45}{2\pi^4}\,\zeta(3)\,\frac{\Gamma_{\nu}^4\,\rho_\gamma(z)}{k_{\rm B} T_{\nu,0}(1+z)}\,c^2 M_{\nu}\equiv\,n_{\nu}(z)\, c^2 M_{\nu}\,,
\label{neutrino-late}
\end{equation}
$n_{\nu}(z)$ being the neutrino number density. 
Massive neutrinos also modify the shape of the matter power spectrum, both in the linear and nonlinear regimes.
First, as neutrinos behave like radiation in the early Universe, they move the radiation-matter equality to slightly later times, thereby shifting the peak of the power spectrum towards smaller wavenumbers.
Second, after the non-relativistic transition, they slow the linear growth of perturbations at scales smaller than the free-streaming length, leading to a scale-dependent growth rate.
The small-scale suppression in the linear power spectra of CDM and baryons, $P_\mathrm{cb}$, or total matter (which includes massive neutrinos), $P_\mathrm{m}$, with respect to a model with massless neutrinos, can be quantified as \citep{Castorina:2015bma}
\begin{equation}
\frac{\Delta P_\mathrm{cb}}{P_\mathrm{cb}}\approx 6\, f_\nu\,,
\qquad
\frac{\Delta P_\mathrm{m}}{P_\mathrm{m}}\approx 8\, f_\nu\,,
\label{eq:pk_supp_cdm_baryons}
\end{equation}
respectively, where $f_\nu \equiv \Omega_\nu / \Omega_\mathrm{m}$ is the neutrino mass fraction.
In the nonlinear regime, this suppression is even more prominent and exhibits a dip at $k\approx 1 \ h \, \mathrm{Mpc}^{-1}$ at low redshifts, giving rise to the well-known `spoon-like’ feature \citep{Hannestad:2020rzl}.

\subsubsection{Primordial non-Gaussianity}

Whereas the most simplistic scenarios of inflation predict that the primordial fluctuations of gravitational potential, $\phi(\vec k)$, follow a distribution that is very close to Gaussian, more complex inflationary models tend to predict more significant deviations from Gaussianity. In the case of a Gaussian random field, all the information is contained in the power spectrum or the 2PCF, with vanishing higher-order correlation functions. Hence, any contribution of primordial 
higher-order correlations is considered PNG. 

The PNG can arise in many forms. Those that arise in the form of primordial bispectrum (or 3-point function) are typically parametrised by $\fnl$. More specifically, one of the most commonly studied PNGs is the so-called local $\fnl$, which transforms a Gaussian  field $\phi_{\rm G}$ into a non-Gaussian field $\phi_{\rm NG}$ as follows

\begin{eqnarray}
    \phi_{\rm NG}(\vec x) =  \phi_{\rm G}(\bm{x}) + \fnl \left[ \phi_{\rm G}(\vec x)^2 - \langle \phi_{\rm G}(\vec x)  ^2\rangle  \right]\,.
\end{eqnarray}
This leads to a non-vanishing primordial bispectrum given by 
\begin{equation}
\begin{split}
 B_{\phi}(\vec k_1, \vec k_2, \vec k_3) = &\fnl \Big[P_{\phi}(k_1)P_{\phi}(k_2) \\
 &+ P_{\phi}(k_2)P_{\phi}(k_3) + P_{\phi}(k_3)P_{\phi}(k_1) \Big]\,.    
\end{split}
\end{equation}
Typically, multi-field models of inflation have a local $\fnl$ of order unity or larger, whereas canonical single-field slow-roll inflation would have a much smaller value, making this parameter a unique signature of primordial physics. 

In the late-time large-scale structure, one could find signatures of this PNG in the bispectrum of matter, as well as in other statistics such as the halo mass function or the power spectrum of galaxies. The latter has become the most promising way to reach the $\sigma(\fnl)\sim 1$ barrier. This arises from an effect known as PNG scale-dependent bias \citep{dalal2008imprints,slosar2008constraints,matarrese2008efect}, which can be expressed as

\begin{eqnarray}
    \delta_{\rm g} (\bm{k}) = \left(b_1 + \frac{b_\phi \,\fnl}{\alpha(k)}\right) \delta_{\rm m} (\bm{k})\,,
\end{eqnarray}
where $k= |\bm{k}|$, and $\alpha(k)$ is defined as
\begin{eqnarray}
\alpha(k) = \frac{2 c^2k^2 T(k)\,D(z)}{3 \Omega_{\rm m,0}\,H_0^2}  \,,
\label{eq:alpha_k}
\end{eqnarray}
with $D(z)$ the normalised linear growth factor.
This way, the PNG introduces a dependency of $1/k^2$ in $\delta(k)$ that can dominate the signal of the galaxy power spectrum at large scales ($k\to 0$). It is among the primary goals of \Euclid to constrain the local PNG by studying this effect. 

\subsubsection{\wwcdm}
A convenient phenomenological description of the dark energy fluid beyond standard \lcdm is the \wwcdm model, where the cosmological constant is replaced by a dynamical, time-dependent dark energy component. It affects the formation of structures through a modification of the expansion history. In a flat geometry ($\Omega_k = 0$), the Hubble parameter $H(z)$ is given by
\begin{equation}
    H(z) = H_0 \sqrt{\Omega_{\rm m,0} (1+z)^3 + \Omega_{\rm r,0} (1+z)^4 + \Omega_{\rm DE,0}f_{\rm DE}(z)}\,,
\end{equation}
with $\Omega_{\rm r}$ the radiation density, accounting for photons and massless neutrinos, while we neglected the effect of massive neutrinos.
Since dark energy is no longer a cosmological constant, its energy density is also not constant, with a redshift dependence given by 
\begin{equation}
    f_{\rm DE}(z) = \frac{\rho_{\rm DE}(z)}{\rho_{\rm DE,0}} =  \exp{\left\{\int_0^z\frac{3\left[1+w(z')\right]\diff z'}{1+z'} \right\}}\,.
\end{equation}
Assuming the popular Chevalier--Polarski--Linder (CPL) parametrisation \citep{chevallier2001accelerating, linder2003exploring}, the dark energy equation of state can be expressed as
\begin{equation}
    w(z) = w_0 + w_a \frac{z}{1+z}\,.
\end{equation}
It is one of the simplest and easiest modifications from \lcdm (that we recover for $w_0 = -1$ and $w_a = 0$), which do not modify the Einstein field equations or the equations of motion.

\subsubsection{Interacting dark energy}

While parametric forms of the dark energy equation of state $w(z)$ -- like the CPL parametrisation discussed above -- provide a first extension of the dark energy phenomenology from the standard cosmological constant, they lack any physical insight into its fundamental nature. A more physically motivated model for the dark energy component sourcing cosmic acceleration can be built by introducing a classical scalar field evolving in a self-interacting potential \citep[][]{Wetterich:1987fm,Ratra:1987rm}
. Such models, known as `Quintessence', feature scaling solutions \citep[][]{Copeland:1997et} that may allow us to mitigate the fine-tuning and coincidence problems arising from a cosmological constant $\Lambda $ \citep[][]{Steinhardt:1999nw}. 

By introducing an additional scalar field $\phi $ (i.e., the Quintessence field), one may need to consider its possible interactions with the other matter-energy components of the Universe. In fact, although a minimally-coupled scalar field (i.e., interacting only through gravity) represents the simplest possible Quintessence scenario, no fundamental principle allows one to exclude a priori other possible direct interactions of the field \citep[][]{Wetterich:1994bg}, whose strength can only be constrained by observations. In particular, a direct coupling of a light scalar field with baryonic matter is already tightly constrained by local gravity experiments \citep[][]{Bertotti:2003rm,Will:2014kxa}. The latter, however, do not constrain a direct interaction with dark matter particles that could leave imprints on both the cosmic-background evolution and the growth of linear and nonlinear structures.

More specifically, `Coupled Quintessence' models \citep[][]{Wetterich:1994bg,Amendola:1999er}, also known as `Interacting dark energy', feature a vanishing coupling between the dark energy scalar field and baryons, while allowing a non-vanishing interaction, usually parametrised through a coupling strength $\beta$ with the dark matter sector, leading to an additional force (equal to a Newtonian force for $\beta = 1/\sqrt{2}$). Such models are fully characterised, at the background level, by a closed system of dynamic equations
\begin{eqnarray}
\label{eq:cq_phi}
    \ddot{\phi } + 3H\dot{\phi } + \frac{\diff V}{\diff\phi } &=& \beta \frac{\rho_{\rm c}}{M_{\rm Pl}}\,,\\
    \label{eq:cq_dm}
    \dot{\rho}_{\rm c} + 3H\rho_{\rm c} &=& -\beta \frac{\rho_{\rm c}\dot{\phi }}{M_{\rm Pl}}\,,\\
    \dot{\rho}_{\rm b} + 3H\rho_{\rm b} = \dot{\rho }_{\rm r} + 4H\rho_{\rm r} &=& 0\,,\\
    3H^{2} &=& \frac{1}{M_{\rm Pl}^{2}}(\rho_{\rm r} + \rho_{\rm b} + \rho_{\rm c} + \rho _{\phi})\,,
\end{eqnarray}
where dots correspond to time derivatives, $V(\phi )$ is the scalar self-interaction potential, $M_{\rm Pl} \equiv 1/\sqrt{8\pi G}$ is the reduced Planck mass, and the subscripts b, c, and r indicate baryonic matter, dark matter, and relativistic particles, respectively. The interaction terms coupling Eqs.~\eqref{eq:cq_phi} and \eqref{eq:cq_dm} are responsible for a modified evolution of the dark energy and dark matter densities, which result in a modified background expansion history. Most noticeably, the dark matter density $\rho _{\rm c}$ does not scale anymore like the inverse of the physical volume of the Universe. This implies a variation of the dark matter particle mass with the evolution of the scalar field, as \citep[][]{Amendola:1999er}
\begin{equation}
\label{eq:cq_mass_variation}
    M_{\rm c}(\phi ) = M_{\rm c,0}\exp\{-\beta (\phi-\phi _{0})/M_{\rm Pl}\}\,,
\end{equation}
where $M_{\rm c}(\phi )$ is the dark matter particle mass (as a function of the scalar field value) and $M_{\rm c,0}=M_{\rm c}(\phi = \phi _{0})$ is its present-day value.
Moving to the evolution of perturbations, the interaction with the scalar field mediates an additional attractive force -- besides gravity -- between dark matter particles, such that the acceleration equation for a test dark matter particle in expanding space reads \citep[][]{Wetterich:1994bg,Amendola:2003wa,Baldi:2008ay}
\begin{equation}
\begin{split}
\label{eq:cq_dm_acceleration}
    \dot{\vec{v}}^{\rm c}_{k} = &-H\left[ 1-\frac{\beta \dot{\phi}}{HM_{\rm Pl}}\right]\vec{v}^{\rm c}_{k}\\
    &- \vec{\nabla }\left[ \sum_{i\in {\rm c}}\frac{G (1+2\beta ^{2}) M_{i}(\phi )}{r_{ik}} +\sum_{j\in {\rm b}}\frac{GM_{j}}{r_{jk}}\right]\,,
\end{split}
\end{equation}
where the two sums range over an ensemble of dark matter and baryonic particles, respectively, and $r_{ik}$, $r_{jk}$ are the physical distances between the particles of each of these ensembles and the test dark matter particle $k$. The absence of any coupling between the scalar field and baryonic particles then leaves the corresponding acceleration equation formally unaffected
\begin{equation}
\label{eq:cq_b_acceleration}
    \dot{\vec{v}}^{\rm b}_{k} = -H \vec{v}^{\rm b}_{k} - \vec{\nabla }\left[ \sum_{i\in c}\frac{G M_{i}(\phi )}{r_{ik}} +\sum_{j\in b}\frac{GM_{j}}{r_{jk}}\right]\,,
\end{equation}
even though the presence of the coupling still implicitly appears in the varying mass of the particles in the dark matter ensemble. Therefore, from Eqs.~\eqref{eq:cq_dm_acceleration} and \eqref{eq:cq_b_acceleration}, it appears clear that baryonic and dark matter particles follow different dynamic equations, thereby explicitly showing the violation of the Weak Equivalence Principle \citep[][]{Damour:1994zq,Amendola:1999er} that characterises Interacting dark energy models. Equations~\eqref{eq:cq_dm_acceleration} and \eqref{eq:cq_b_acceleration} have been implemented in the $N$-body codes \citep[][]{Baldi:2008ay,Baldi:2010vv, zhang2018fully} employed for the simulations of Interacting dark energy models discussed below, along with the corresponding modified evolution of the cosmic expansion history $H(z)$ and the variation of dark matter particle masses $M_{\rm c}(\phi)$ in Eq.~\eqref{eq:cq_mass_variation}. 

Besides the energy exchange described by Eqs.~\eqref{eq:cq_phi} and \eqref{eq:cq_dm}, other possible forms of interaction between an evolving dark energy component and dark matter have been proposed in the literature \citep[][]{Pourtsidou:2013nha,Skordis:2015yra}. A particular class of such models, involving pure momentum exchange rather than rest-energy exchange, has attracted significant attention in the last years \citep[][]{Pourtsidou:2016ico,Bose:2017jjx,Carrilho:2021rqo} as it allows modifying the evolution of perturbations without affecting the background cosmic expansion. The most popular example of these momentum-exchange models is the `dark scattering' scenario \citep[][]{Simpson:2010vh}, where dark matter particles moving in a homogeneous dark energy field with equation of state $w$ experience a velocity-dependent force similar to the case of Thomson scattering for electrons moving in a photon fluid. Such force can be parametrised through a cross section $\sigma _{D}$ such that the resulting acceleration for a dark matter particle $k$ in expanding space reads \citep[][]{baldi2016structure}
\begin{equation}
\label{eq:scattering_acceleration}
    \dot{\vec{v}}_{k} = -(1+A)H\vec{v}_{k} + \sum_{j\neq k}\frac{G M_{j}\vec{r}_{jk}}{r_{jk}^{3}}\,,
\end{equation}
where the scattering factor $A$ can be written as
\begin{equation}
\label{eq:scattering_factor}
    A\equiv (1+w)\sigma_{D}\frac{c}{m_{\rm CDM}}\frac{3\Omega_{\rm DE,0}}{8\pi G}H\,,
\end{equation}
with $c$ the speed of light and $m_{\rm CDM}$ the mass of the fundamental dark matter particle. As we have no clear bounds on the mass of dark matter particles,  Eq.~\eqref{eq:scattering_factor} implies that the scattering cross section $\sigma _{D}$ and the dark matter particle mass are degenerate in the determination of the overall scattering force, that will be specified by their ratio. For this reason, dark scattering models are usually parametrised through the scattering strength $\varpi \equiv c\sigma _{D}/m_{\rm CDM}$. Equation~\eqref{eq:scattering_acceleration}, which has been implemented in the $N$-body codes employed for the dark scattering simulations discussed below \citep[see, e.g.,][for the details of the implementation]{Baldi:2014ica,baldi2016structure}, shows that in these models the gravitational interaction between massive particles is not affected by the presence of the new scattering force, which couples to particles' peculiar velocities and whose strength is modulated by the factor $(1+w)$ such that for the limiting case of a cosmological constant ($w=-1$), the scattering force would vanish. This also shows that for phantom-like dark energy models (i.e. models with $w < -1$) the extra force is actually a drag, accelerating particles along their direction of motion.

It is important to stress that, for both Coupled Quintessence  and dark scattering scenarios, a proper modelling of the evolution of perturbation and the consequent formation of nonlinear structures requires taking into account the different dynamics of baryons and dark matter. This implies including two distinct families of numerical particles in the simulations of these cosmologies, as done in the simulations discussed in Sect.~\ref{sec:cider_dakar_simulations}.

\subsubsection{$f(R)$ gravity}

In GR, the Lagrangian of the Einstein--Hilbert action is proportional to $R$ the Ricci scalar (or curvature). In the $f(R)$ gravity theory \citep{buchdal1970nonlinear,sotiriou2010reviews}, we instead allow for an arbitrary function of the curvature. The total action is then 
\begin{equation}
    S = \int \diff^4x \sqrt{-g} \left[ \frac{R + f(R)}{16\pi G} + \mathcal{L}_{\rm m}\right]\,,
\end{equation}
where $\mathcal{L}_{\rm m}$ is the matter Lagrangian, $g$ the determinant of the metric, and $f(R)$ a function of curvature that is equal to $-2\Lambda$ in \lcdm.
A popular parametrization for this model is that of \cite{hu2007models}, which follows the functional form
\begin{equation}
    f(R) = -m^2\frac{c_1\left(R/m^2\right)^n}{c_2\left(R/m^2\right)^n + 1}\,,
\end{equation}
where $n$, $c_1$, and $c_2$ are the parameters of the model, and $m$ is the curvature scale given by $m^2 = \Omegam H_0^2/c^2 = 8\pi G\rho_{\rm m,0}/(3c^2)$, with $\rho_{\rm m,0}$ being the mean matter density today. This model introduces a `chameleon' screening mechanism \citep{khoury2004chameleon,burrage2018tests} to suppress the fifth force arising from the additional scalar field (also called `scalaron'), allowing us to recover GR in high-density regions and therefore pass Solar System experiments. 
The observational evidence of a dark energy component in the form of a cosmological constant imposes the constraint
\begin{equation}
    \frac{c_1}{c_2} = 6\frac{\Omega_{\Lambda, 0}}{\Omega_{\rm m, 0}}\,, \quad {\rm and} \quad \frac{c_1}{c_2^2} = -\frac{1}{n} \left(3+12\frac{\Omega_{\Lambda, 0}}{\Omega_{\rm m, 0}} \right)^{n+1} f_{R0}\,,
\end{equation}
with $f_{R0}$ the background value of the scalaron field $f_R$ today. From now on, we only consider the case of $n = 1$, leaving $f_{R0}$ as the only free parameter that drives the modifications from GR, which is constrained to $\logten |f_{R0}| \lesssim -5.$ from galaxy cluster measurements \citep{koyama2016cosmological,vogt2025constraints}. In practice, the Poisson equation is modified compared to the Newtonian case, with an additional source term that depends on the scalaron field
\begin{eqnarray}
    \nabla^2\Phi &=& 4\pi Ga^2 \delta\rho - \frac{1}{2}c^2\nabla^2 f_R, \\
    \nabla^2 f_R &=& -\frac{8\pi Ga^2}{3c^2} \delta\rho \nonumber \\
        &&+ \frac{m^2 a^2}{3}  \left( \sqrt{-\frac{1}{f_R} \frac{c_1}{c_2^2}} + 3a^{-3} + 12\frac{\Omega_{\Lambda, 0}}{\Omega_{\rm m, 0}} \right)\,.
\end{eqnarray}
We refer the interested reader to \citet{EP-Adamek}, and references therein, for a comprehensive code comparison project on numerical methods to run simulations in \cite{hu2007models} $f(R)$ gravity.
The main observational feature of $f(R)$ gravity compared to \lcdm is that, although they share the same expansion history, $f(R)$ has higher small-scale clustering, with an amplitude and shape that depend on $f_{R0}$.

\subsection{The \vdg galaxy-clustering model}
\label{sec:vdg_theory}
To analyse the two-point statistics measured in the simulations, we employ the state-of-the-art velocity difference generator model \cite[hereafter, \vdg model,][]{sanchez2017clustering}.
The model is implemented in \comet (see \citealt{eggemeier2023comet} and \citealt{pezzotta2025extending}, the latter being the recent extension to include massive neutrinos), a fast emulator for perturbation theory based on Gaussian processes that allows for an efficient analysis of the posterior distribution of the main cosmological parameters.
It extends the original formulation of the \vdg model to include renormalisation of loop contributions via the effective field theory of LSS (hereafter, EFT, \citealt{BaumannEtal2012}, \citealt{carrasco2012effective}).
In addition, it includes full infrared-resummation based on the wiggle-no-wiggle split of the power spectrum \citep{Blas2016a, Baldauf-ir-res2015}.
At its base, the model builds on the following expression for the power spectrum \citep{scoccimarro2004redshift, taruya2010baryon} 
\begin{equation}
\begin{split}\label{eq:Pk_redshiftspace}
    P^{s}(\kv) = \int &\diff^3 r\, \mathrm{e}^{\mathrm{i}\kv\cdot\bm{r}} \, \Big\langle \exp\left\{-\mathrm{i}k \mu f \Delta u_z\right\}\\
    &  \brackets{\delta(\xv)+f\,\nabla_z u_z(\xv)}\,\brackets{\delta(\xv') + f\, \nabla_z u_z(\xv')}\Big\rangle\,,
\end{split}
\end{equation}
with the rescaled velocity $u_z(\xv) \equiv -v_z\,/\,(aHf)$ into the $z$-direction (plane parallel approximation) and its difference $\Delta u_z \equiv u_z(\bm{x})-u_z(\xv')$.
Moreover, $\mu$ is defined via $\kv\cdot\hat{\bm{z}} = k\,\mu$, the pair-separation vector is given by $\bm{r} = \xv-\xv'$, and $f$ is the growth rate.
This expression can be derived from an expansion of the Jacobian coming from the real-to-redshift space mapping.
Using the cumulant-expansion theorem, this moment can be further massaged into a set of cumulants involving both the density and velocity fields with a common pre-factor consisting of an exponentiated cumulant of two velocity fields that is hence called the velocity difference generator.
\citet{taruya2010baryon} approximated this generator in their model called `TNS', with either a Lorentzian or a Gaussian function showing an improved modelling of the small-scale Fingers-of-God effect \citep{Jackson1972MNRAS}.
The TNS model has been further improved by \cite{sanchez2017clustering} with the proposition of a more complex form for the damping function taking the form of a Gaussian with kurtosis motivated by the large-scale limit of the PDF of pairwise velocities \citep{scoccimarro2004redshift}.
The damping function reads
\begin{equation}\label{eq:VDG_damping_function}
    D(k, \mu, a_{\rm vir}) = \frac{1}{\sqrt{1-\lambda^2a_{\textrm{vir}}^2}}\exp\left (\frac{\lambda^2\sigma_v^2}{1-\lambda^2a_{\textrm{vir}}^2}\right) \,,
\end{equation}
where $a_{\mathrm{vir}}$ is a free parameter taking into account deviations from Gaussianity, $\lambda = \mathrm{i}fk\mu$, and $\sigma_v$ is the linear velocity dispersion.
The EFT treatment of biased clustering in redshift space leads to three counterterms that add to the free parameters of the model.
Galaxy bias is consistently accounted for in the \vdg model with a basis of four renormalised bias parameters \citep{Assassi2014JCAP} and the stochastic contributions to the power spectrum take into account deviations from local Poisson shot noise but also include non-local noise contributions \citep[see Eqs. 20--24 in][for details of the noise and counterterm contributions]{eggemeier2023comet}.

\section{\label{sec:simulations}Simulations}
We consider simulations that have been run with the non-standard models described in Sect.~\ref{sec:non-standard_models}, as well as some reference \lcdm simulations for null testing. More details regarding these simulations can be found in \cite{EP-Racz}, and throughout this paper, we focus on snapshots at $z = 1$. The simulation suites are summarised in Table~\ref{tab:sims}.

\begin{table*}[]
    \centering
     \caption{Overview of the simulation suites analysed for this project. The cosmological parameters for each simulation can be found in Sect.~\ref{sec:simulations}.}
    \begin{tabular}{|c c c c c c| c|}
  \hline
  \rowcolor{blue!5}
  & & & & & & \\[-8pt]
  \rowcolor{blue!5}
 Name & $N_{\rm sim}$ & $L_\mathrm{box}$ [$\gpcoh$] &$N_\mathrm{DM}$ & $m_\mathrm{DM}$ [$\msoh$] & Model &  Main reference
 \\  \hline 
  & & & & & & \\[-8pt]
      \multirow{2}{*}{\RAYGAL} & 1 & $2.625$ & $4096^3$ & $1.88\times10^{10}$ & \lcdm &  \multirow{2}{*}{\cite{rasera2022raygal}}\\ 
    & 1 & $2.625$ &  $4096^3$ & $2.0\times10^{10}$ & \wcdm & 
    \\ \hline
    & & & & & & \\[-8pt]
    \multirow{2}{*}{\COMPLEMENTARY}  & $2$ & $1.5$ & $2160^3$ & $2.89\times10^{10}$ & \lcdm & \multirow{2}{*}{\cite{racz2023complementary}} \\
    & $2$ & $1.5$ & $2160^3$ & $2.88\times10^{10}$ & \wcdm &  
    \\ \hline
    & & & & & & \\[-8pt]
    \PNGUNITsim & 1 & $1.0$ & $4096^3$ & $1.2\times10^9$ & PNG &  \cite{adame2024PNG}
    \\ \hline 
    & & & & & & \\[-8pt]
    \DEMNUni  & $15$ & $2.0$ & $2048^3$ & $8\times10^{10}$ & \wwcdm + $m_{\nu}$ & \cite{DEMNUni_simulations}  
    \\ \hline
    & & & & & & \\[-8pt]
    \CIDER  & $4$ & $1.0$ & $1024^3$ & $8.1\times10^{10}$ & Clustering DE &  \cite{baldi2023cider} 
    \\ \hline
    & & & & & & \\[-8pt]
    \DAKARTWO & $5$ & $1.0$ & $1024^3$ & $8.1\times10^{10}$ & Dark Scattering &  \cite{baldi2016structure} 
    \\ \hline
    & & & & & & \\[-8pt]
    \multirow{2}{*}{\DUSTGRAINPATHFINDER} & 1 & \multirow{2}{*}{0.75} & \multirow{2}{*}{$768^3$} & \multirow{2}{*}{$8.1\times10^{10}$} & \lcdm &  \multirow{2}{*}{\cite{giocoli2018weak}} \\
     & 1 &  &  &  & $f(R)$ &  \\
    \hline
    \end{tabular}
    \label{tab:sims}
\end{table*}
\subsection{The \FLAGSHIPTWO simulation}
\label{sec:fs2}

\begin{table}
\small
    \caption{Cosmological parameters of the Flagship 2 simulation \citep{EuclidSkyFlagship}.}
  \smallskip
  \label{tab:flagship_cosmology}
  \smallskip
  \begin{tabular}{c c c c c c c}
    \hline
    \rowcolor{blue!5}
    & & & & & &  \\[-8pt]
    \rowcolor{blue!5}
    $h$ & $\Omega_{\rm m,0}$ & $\Omega_{\rm b,0}$ & $10^4\,\Omega_{\rm r,0}$ & $M_\nu\,[\rm{eV}]$ & $10^9\,A_{\mathrm{s}}$ & $n_{\rm s}$ 
    \\ \hline 
    & & & & & &  \\[-8pt]
    0.67 & 0.319 & 0.049& 0.5509 & 0.0587 &  2.1 & 0.96 \\
    \hline
  \end{tabular}
   \tablefoot{The table shows the value of the Hubble parameter $h$, the total matter density $\Omega_{\rm m,0}$, baryon density $\Omega_{\rm b,0}$, radiation density $\Omega_{\rm r,0}$, the total neutrino mass $M_\nu$, the scalar amplitude $A_{\rm s}$, and the spectral index $n_{\rm s}$.}
    \label{tab:fs2cosmo}
\end{table}


The reference simulated galaxy catalogue for the Euclid Wide Survey is the so-called `Flagship 2' simulation
\citep[][hereafter FS2]{EuclidSkyFlagship}. The FS2 features a simulation box of $3600\mpcoh$ on each side with $16\,000^3$ particles, leading to a mass resolution of $m_{\rm p} = 10^9\,h^{-1}M_{\odot}$. This 4 trillion particle simulation is the largest $N$-body simulation performed to date and matches the basic science requirements of the mission because it allows us to include the faintest galaxies that \Euclid will observe while sampling a cosmological volume comparable to what the satellite will survey, although with lower field of view (one octant of the sky, meaning roughly 5100\,deg$^2$, while \Euclid\ should reach 14 000\,deg$^2$ by the end of the mission). 

The simulation was performed using \texttt{PKDGRAV3} \citep{Potter:16} on the Piz Daint supercomputer at the Swiss National Supercomputer Center (CSCS).
The input cosmology used in the simulation is given in Table \ref{tab:fs2cosmo}, whereas the initial conditions were realised at $z = 99$ with first-order Lagrangian perturbation theory (1LPT) displacements from a uniform particle grid (for further details, see \citealt{EuclidSkyFlagship}).  The main data product was produced on the fly during the simulation and is a continuous full-sky particle light cone out to $z=3$,  which was used to identify about 150 billion dark matter haloes with \rockstar.
The halo catalogue (along with the set of two-dimensional dark matter maps also produced on the fly) is the main input for the FS2 galaxy catalogue. Galaxies were generated following a combination of HOD and abundance matching (AM) techniques. Following the HOD prescription, haloes were populated with central and satellite galaxies. Each halo contains a central galaxy and a number of satellites that depend on the halo mass.  The halo occupation was chosen to reproduce observational constraints of galaxy clustering in the local Universe \citep{Zehavi2011}. 
A detailed description of the catalogue production and implementation of galaxy properties is given in \cite{EuclidSkyFlagship}.

\subsection{The \RAYGAL simulations}

The \RAYGAL simulations, as detailed by \cite{breton2019imprints} and \cite{rasera2022raygal}, consist of two dark matter only simulations run with the $N$-body code \RAMSES \citep{teyssier2002cosmological}. These simulations have been run with a WMAP-7yr cosmology \citep{komatsu2011seven}, that is $h = 0.72$, $n_{\rm s} = 0.963$, $\Omega_{\rm b,0} = 0.04356$, and $\Omega_{\rm r,0} = 0.04356$. For the $\Lambda$CDM simulation we have $\Omega_{\rm m,0} = 0.25733$ and $\sigma_8 = 0.80101$, while the \wcdm simulation has $\Omega_{\rm m,0} = 0.27508$, $\sigma_8 = 0.85205$, and $w = -1.2$. The initial condition module employs second-order Lagrangian perturbation theory (2LPT) to mitigate transient effects \citep{crocce2006transients} at $z_{\rm ini} = 46$.

\subsection{The \COMPLEMENTARY simulations}
The \COMPLEMENTARY simulation suite consists of two pairs of dark matter only simulations in \lcdm and \wcdm cosmologies. Each set of simulations is started from a phase-shifted and matched amplitude initial condition pair as described in \cite{racz2023complementary}. One of each pair is called `original' simulation, and it is started from regular Gaussian initial conditions. The second simulation in the pair is called `complementary', and the initial amplitudes of this run were modified to ensure that the average power spectrum of the pair is equal to the cosmic mean power spectrum from linear theory. The averaged mock observables calculated from such a pair have suppressed cosmic variance, while one of the simulations is always started from regular Gaussian initial conditions. The initial conditions were generated by using 1LPT with the \NGENIC code \citep{ngenic2015} at redshift $z_{\rm ini}=127$, and the initial transfer function was calculated with the \CAMB Boltzmann solver \citep{camb2011}. The simulations were run with the \GIZMO \citep{gizmo2015} code. The cosmological parameters of the simulations in this work are consistent with the {\it Planck}-2018 \citep{planck2018cosmological} observations. In the \lcdm pair, the parameters were $\Omega_{\rm m, 0}=1-\Omega_{\rm DE,0}=0.3111$, $\Omega_{\rm b,0}=0.04897$, $h=0.6766$, $n_{\rm s}=0.9665$, and $\sigma_8=0.8102$. The \wcdm pair used the same Gaussian white noise in the initial condition generation, and the cosmological parameters were set to $w_0=-1.04$, $\Omega_{\rm m, 0}=0.3096$, $\Omega_{\rm b, 0}=0.04899$, $\Omega_{\rm DE,0}=0.6904$, $h=0.6766$, $n_{\rm s}=0.9331$, and $\sigma_8=0.8438$.

\subsection{The \PNGUNITsim simulation}

The \PNGUNITsim simulation \citep{adame2024PNG} is a twin of one of the original (Gaussian) \UNITsims \citep{chuang2019unit}, but with local PNG given by $\fnl=100$. The parameters are based on a {\it Planck} \lcdm cosmology with $\Omega_{\rm m,0} = 1 - \Omega_{\Lambda,0} = 0.3089, \hspace{0.25cm} h = 0.6774, \hspace{0.25cm} n_{\rm s} = 0.9667, \hspace{0.25cm} \sigma_{8} = 0.8147$. This simulation set its initial conditions at $z=99$ with the \FASTPM code using $4096^3$ particles in a $\left( 1 h^{-1}{\rm Gpc}\right)^3$ volume with 1LPT. The particles are evolved with the tree-PM code \LGADGETTWO, which is a non-public version of \GADGETTWO optimised for heavy memory use runs.\footnote{This feature has now been incorporated into the public version of \GADGETFOUR \citep{springel2021simulating}}.
One particularity of all the \UNITsims is that they use fixed initial conditions. This means that the modes of the initial conditions are set to their expectation value, significantly reducing the variance of several statistics \citep{angulo2016cosmological}, making it ideal for testing models of galaxy formation. The usage of this technique for local PNG was validated in \cite{avila2023validating} and was also implemented in the \PNGUNITsim. Additionally, the phases of the \PNGUNITsim initial conditions are identical to one of the four \UNITsims. This is what is referred to as `matched' simulations in \cite{avila2023validating}.

\subsection{The \DEMNUni simulations}
The `Dark Energy and Massive Neutrino Universe' (\DEMNUni) simulations~\citep{DEMNUni_simulations} have been produced with the aim of investigating large-scale structures in the presence of massive neutrinos and dynamical dark energy, and were conceived for the nonlinear analysis and modelling of different probes, including dark matter, halo, and galaxy clustering \citep[see][Carella et al., in prep.]{Castorina:2015bma, Bel2019, Parimbelli2021, Gouyou_Beauchamps_2023}, weak lensing, CMB lensing, SZ and ISW effects \citep{DEMNUni_simulations, EP-Ingoglia, Luchina_etal_2025}, cosmic void statistics \citep{Kreisch2019, Schuster2019, verza_2019}, and cross-correlations among these probes \citep{Vielzeuf2023, Cuozzo_2024}.
The \DEMNUni simulations are characterised by a softening length $\varepsilon=20\, h^{-1} \, {\rm kpc}$, a comoving volume of $8 \, h^{-3} \, \mathrm{Gpc}^3$ filled with $2048^3$ dark matter particles and, when present, $2048^3$ neutrino particles. The simulations are initialised at $z_{\rm ini}=99$ with 1LPT. The initial power spectrum is rescaled to the initial redshift via the rescaling method developed in~\cite{zennaro_2017}. Initial conditions are then generated with a modified version of the \texttt{N-GenIC} software.
The \DEMNUni simulations were run using the tree particle mesh-smoothed particle hydrodynamics (TreePM-SPH) code \texttt{P-Gadget3} \citep{springel2005cosmological}, specifically modified as in \cite{Viel:2010bn} to account for the presence of massive neutrinos. This modified version of \texttt{P-Gadget3} follows the evolution of CDM and neutrino particles, treating them as two separate collisionless components. 
The reference cosmological parameters are chosen to be close to the baseline {\it Planck} 2013 cosmology \citep{planck2014overview}
$$\{\Omega_{\rm b,0}, \Omega_{\rm m,0}, h, n_{\rm s}, A_{\rm s} \} = \{0.05, 0.32, 0.67, 0.96, 2.127 \times 10^{-9} \}.$$
Given these values, the reference (i.e., the massless neutrino case) CDM-particle mass resolution is $m^{\rm p}_{\rm CDM} = 8.27\times 10^{10} \, h^{-1} \, \si{\solarmass}$ and decreases according to the mass of neutrino particles in order to maintain the same $\Omega_{\rm m,0}$ among all the \DEMNUni simulations. In fact, massive neutrinos are assumed to be a particle component in a three mass-degenerate scenario. To keep $\Omega_{\rm m,0}$ fixed, an increase in the massive neutrino density fraction yields a decrease in the CDM density fraction.

\subsection{The \CIDER/\DAKARTWO simulations}
\label{sec:cider_dakar_simulations}
The \CIDER \citep{baldi2023cider} and \DAKARTWO \citep[see, e.g.,][]{EP-Racz} simulations are a suite of cosmological dark matter only simulations for two different interacting DE scenarios, namely the `Constrained Interacting Dark Energy' \citep[for \CIDER,][]{Barros:2018efl} and the dark scattering \citep[for \DAKAR,][]{Simpson:2010vh} models, respectively.
These different simulation suites share the same reference \lcdm cosmology as a limiting case, with cosmological parameters $\Omega _{\rm m,0} = 0.262$, $\Omega _{\rm b,0}=0.049$, $\Omega _{\Lambda,0 } = 0.689$, $\Omega_{k,0} = 0$, $h=0.677$, $n_{\rm s}=0.9665$, and $A_{\rm s}=1.992 \times 10^{-9}$. The simulations follow the evolution of $2\times 1024^3$ particles for dark matter and baryons in a box of $1\gpcoh$ per side, where the baryonic particles are treated as a separate collisionless matter component (i.e., neglecting hydrodynamical forces) in order to properly model the gravitational effects of the DE-CDM interaction. Initial conditions have been generated at $z_{\rm ini}=99$ by means of the public code \texttt{MUSIC} \citep[][]{hahn2013music} adopting 2LPT. Both simulation suites have been run with the \texttt{C-Gadget} code \citep[][]{Baldi:2008ay}, a modified version of the \texttt{Gadget-3} code that covers a wide range of dark energy models.

Having the same reference \lcdm realisation as a limiting case, the \CIDER simulations  include three additional runs for different values of the DE-CDM coupling constant $\beta \in \left\{0.03, 0.05, 0.08 \right\}$, while sharing the same background expansion history as the \lcdm cosmology. This is the main feature of this class of interacting DE models, where the cosmic background evolution is identical to \lcdm by construction, while the growth of perturbations will be modified by the presence of the interaction, giving rise to a different structure formation history \citep[for more details, see][]{Barros:2018efl,baldi2023cider}.
The \DAKAR simulations, instead, do not share the same background expansion history as their \lcdm reference. In fact, in this class of models, the scattering force acting on the interacting DM particles vanishes for a DE equation of state parameter $w=-1$. The \DAKARTWO simulations, therefore, explored different possible behaviours of a varying equation of state parameter following the CPL parametrisation $w(a) = w_{0} + (1-a)w_{a}$, including one `freezing' scenario with $(w_{0},w_{a})=(-0.95,+0.1)$, two `thawing' models departing from $w=-1$ at low redshifts in the Quintessence $w>-1$ and in the `phantom' $w<-1$ regimes, with parameters $(w_{0},w_{a})=(-0.9,-0.1)$ and $(w_{0},w_{a})=(-1.1,+0.1)$, respectively, and a `phantom crossing' model with $(w_{0},w_{a})=(-0.9,-0.2)$. For all these models, the scattering parameter $\varpi$ has been set to 100 bn/GeV.

\subsection{The \DUSTGRAINPATHFINDER simulations}

The \DUSTGRAINPATHFINDER simulations \citep[][]{giocoli2018weak} are a suite of cosmological dark matter only simulations aimed at sampling the $2$-dimensional $(\Sigma M_{\nu}, f_{R0})$ parameter space for cosmological models characterised by the simultaneous presence of a non-vanishing total neutrino mass ($\Sigma M_{\nu} \in [0.0,0.3]$ eV) and a non-standard theory of gravity, in particular considering the \citet{hu2007models}  $f(R)$ gravity with $|f_{R0}| \in [0.0, 10^{-4}]$. In fact, it is well known that the effects of massive neutrinos and of $f(R)$ modified gravity on cosmological observables are strongly degenerate \citep[][]{Baldi:2013iza}.
The \DUSTGRAINPATHFINDER simulations follow the evolution of $768^3$ particles in a cosmological comoving volume of $750\mpcoh$ per side, including different possible combinations of the $\Sigma M_{\nu}$ and $f_{R0}$ parameters, for a total of 10 simulations, while keeping all other standard cosmological parameters fixed to their reference \lcdm values: $\Omega _{\rm m,0}\equiv \Omega _{\rm cb,0} + \Omega _{\rm b,0} + \Omega _{\nu,0} = 0.31345$, $\Omega _{\rm b,0} = 0.0481$, $\Omega _{\Lambda,0 } = 0.68665 $, $\Omega _{k} = 0$, $h=0.6731$, $n_{\rm s}=0.9658$, $A_{\rm s}=2.199\times 10^{-9}$ which results in a value of $\sigma _{8}=0.842$ for the reference \lcdm model. An extension of the original \DUSTGRAINPATHFINDER simulations including ten additional  runs for standard gravity and for a \wcdm cosmology by varying standard cosmological parameters was recently employed by \citet[][]{Euclid:2023uha} to forecast \Euclid constraining power using higher-order weak lensing statistics. All the simulations of the \DUSTGRAINPATHFINDER series have been run with the \texttt{MG-Gadget} code \citep[][]{puchwein2013mggadget}, a modified version of the \texttt{Gadget-3} code for $f(R)$ -- while the neutrino component has been treated using the particle method described in \citet[][]{Viel:2010bn}, see also \cite{Euclid:2022qde} for a comparison of different numerical methods to model massive neutrinos.

\section{\label{sec:Methods}Methods}

We process all the simulations through a common end-to-end pipeline, which takes as inputs particle snapshots and outputs galaxy catalogues and 2PCF  multipoles in redshift space.

\subsection{Halo catalogue production}

In this section, we summarize the main properties of our halo catalogue pipeline that are useful for our paper. More details can be found in \cite{EP-Racz}.
We ran the \rockstar halo finder on every simulation snapshot presented in Sect.~\ref{sec:simulations}. \rockstar makes a first selection employing a three-dimensional friends-of-friends (FoF) algorithm  with a linking length $b = 0.28$ to detect haloes with at least ten particles. The halo finder then refines its detection by using FoF in phase space, thus discarding unbound particles: such an estimation of the mass is called `bound mass'. If the estimation contains also unbound particles, then it is called spherical overdensity (SO) masses. \rockstar provides us a halo catalogue containing various information such as position, velocity, SO, and bound mass at various overdensity thresholds. Although FS2 uses haloes with bound \mvir, we chose instead the SO \Mvirm\, definition for all the simulations considered in the present paper. This decision was forced by the requirement to be agnostic to the theory of gravity, which is not the case for bound masses nor \mvir. The closest mass definition was thus \Mvirm, since $\rho_{\rm vir} \approx \rho_{200b} = 200\rho_{\rm m}$ at $z = 0$. Notably, in our case the mass-definition discrepancy between FS2 and the other simulations is not an issue since  we do not resort to any abundance matching technique. In fact, within the HOD framework, any mass rescaling can be absorbed in a re-definition of the parameters of the model.

\subsection{Galaxy catalogue production}
\label{sec:galaxy_catalogue_production}

To emulate galaxies in dark matter simulations we use the HOD framework \citep{berlind2003hod, zheng2005theoretical}. The idea is to compute the conditional probability of having central and satellite galaxies in dark matter haloes, as a function of the halo mass. This framework, fitted on full hydrodynamical simulations (which require heavy computational resources and are usually restricted to the \lcdm model), allows us to mimic the properties of galaxies (such as positions and velocities) at a lower computational cost. This model has also been extended to account for properties beyond halo mass, also known as `assembly bias' \citep{sheth2004environmental, gao2005age}. 

In this work, however, we focus on a simple, although flexible mass-dependent-only HOD model similar to that presented in \cite{avila2020sdss}, in which the mean number of central galaxies per halo as a function of the halo mass $M$ is
\begin{equation}
\ev{\ncen(M)} =
\left\lbrace
\begin{array}{lc}
A_{\rm c}\exp\left(-\frac{\left\{\logten [M/(h^{-1}\si{\solarmass})] - \mathcal{M}\right\}^2}{2\sigma^2}\right),   & \frac{M}{(h^{-1}\si{\solarmass})} < 10^{\mathcal{M}}\,,  \\
A_{\rm c}\left(\frac{M}{10^\mathcal{M}}\right)^\gamma,  & \frac{M}{(h^{-1}\si{\solarmass})} \geq 10^{\mathcal{M}}\,,
\end{array}\right.
\label{eq:ncen}
\end{equation}
where $A_{\rm c}$ is the mean number of centrals at $\logten \left[M/(h^{-1}\si{\solarmass})\right] = \mathcal{M}$, and $10^\mathcal{M}\,h^{-1}\si{\solarmass}$ is the mass at which we transition between a log-normal  with variance $\sigma^2$ to a power-law with exponent $\gamma$. We then sample $\ncen$ with a Bernoulli distribution, placing the central galaxy at the halo centre with the same peculiar velocity. 

The shape of this central HOD function is characteristic of emission-line galaxies (ELGs) and was fitted in \citet{avila2020sdss} to a semi-analytical model of galaxy formation presented in \citet{gonzalez-perez2018host}. Similar shapes were found by other studies when examining ELGs \citep{zhai2021linear,reyes-peraza2024improved}. This shape is different from the traditional central HOD curve given by a (smoothed) step function \citep{zheng2005theoretical}, which is more suited for mass or magnitude-limited samples.

The satellite occupation is given by
\begin{equation}
\ev{\nsat} = (1-\ev{\ncen})\ev{\nsat | 0} + \ev{\ncen}\ev{\nsat | 1} \,,
\end{equation}
with $\ev{\nsat | \ncen}$ the mean number of satellites in haloes that contain $\ncen$ central galaxies. Such a distinction between the statistical occupation of satellite galaxies, depending on the presence of a central galaxy, is also called `conformity' \citep{rocher2023desi,gao2024desi}. In this paper, we consider a simple conformity relation that differs only in the overall amplitude. Note that in \cite{reyes-peraza2024improved}, a similar implementation was shown to work as well as a more complex mass-dependent conformity. Finally, its functional form is given by
\begin{equation}
\ev{\nsat(M) | N} = A_{{\rm sat},N}\left(\frac{M - M_0}{M_1}\right)^\alpha\,,
\label{eq:nsat}
\end{equation}
with $M_0$ the minimum halo mass which can contain a satellite and $M_1$ the typical halo mass at which we have $A_{{\rm sat},N}$ satellites per halo (in the limit $M_0 \ll M_1$) and $\alpha$ controls the slope of the number of satellites per halo in the high-mass end. In the case of satellites, the power law is more standard for all types of tracers.

We fix the mean number density
\begin{equation}
    \bar{n}_{\rm gal} = \int \frac{\diff n(M)}{\diff M} \left[\ev{\ncen(M)} + \ev{\nsat(M)}\right] \diff M\,,
\end{equation}
to that of H$\alpha$ galaxies $\bar{n}_{\mathrm{H}\alpha}$ given by the model 3 of \cite{pozzetti2016modelling}.
We fix $A_{\rm c} = {\rm min}\left[\bar{n}_{\mathrm{H}\alpha}/\bar{n}_{\rm gal}, {\rm max}\left(\ev{\ncen}\right)^{-1}\right]$, while at large scales the linear bias is given by
\begin{equation}
    b_{\rm gal} = \frac{1}{\bar{n}_{\rm gal}}\int \frac{\diff n(M)}{\diff M} b(M) \left[\ev{\ncen(M)} + \ev{\nsat(M)}\right] \diff M\,,
\end{equation}
with $b(M)$ the halo bias as a function of mass.
To fix the mean number density while keeping the same large-scale clustering, we need to rescale $A_{\rm sat,0}$ as
\begin{equation}
    A_{\rm sat, 0} \xrightarrow{} A_{\rm sat,0} \frac{A_{\rm c}\left[1 - \ev{\ncen}\right]}{1 - A_{\rm c}\ev{\ncen}}\,.
\end{equation}
For simplicity, we sample $\nsat$ from a Poissonian distribution, although there exist other possibilities for more complex sampling distributions \citep{vos-gines2024improving}. 
We use a triaxial NFW profiles \citep{navarro1997universal} using halo inertia tensor computed internally in \rockstar and outputted in our modified version. The number density of haloes as a function of radius is 
\begin{equation}
    \rho_{\rm NFW}(r) \propto \left(\frac{r}{R_{\rm s}}\right)^{-1} \left(1+\frac{r}{R_{\rm s}}\right)^{-2}\,,
\end{equation}
with $R_{\rm s}$ the scale radius. In practice, we use an analytical expression for the NFW cumulative probability density from which we can directly draw random radii from a uniform distribution \citep{robotham2018short}. Our implementation is based on that of \cite{EuclidSkyFlagship}, which is an update over the pipeline developed in \cite{carretero2015algorithm} for the MICE simulations \citep{fosalba2015miceA,fosalba2015miceB}.
We also explore the possibility of assigning satellites according to the density profile of H$\alpha$ emitters as proposed by \cite{reyes-peraza2024improved} and which yields
\begin{equation}
    \rho_{{\rm H}\alpha}(r) \propto \left(\frac{r}{r_0}\right)^{\eta-2} \left[1+\left(\frac{r}{r_0}\right)^\beta\right]^\kappa\,,
\end{equation}
with $r_0=0.34 \mpcoh$, $\eta=1.23$, $\beta=3.19$, and $\kappa=-2.1$ when fitted to a sample at $z = 1.3$, which is close to the target sample in \Euclid. In this case, we assume that all haloes share the same characteristic scale $r_0$ , as motivated by the findings in \citet{reyes-peraza2024improved}, while for NFW profiles, we use the scale radius $R_{\rm s}$. We assume that modified gravity simulations also produce haloes with such profiles, however with different parameters (for example, haloes in $f(R)$ or nDGP simulations follow different mass-concentration relations compared to $\Lambda$CDM, \citealt{ruan2024emulator}).
Finally, the satellite galaxy velocity is that of the halo's plus a component drawn from an isotropic Gaussian distribution with zero mean and $\sigma_v$ (halo velocity dispersion estimated from the dark matter particles it contains) variance. 

\begin{figure}[htbp!]
\centering
\includegraphics[width=1.0\hsize]{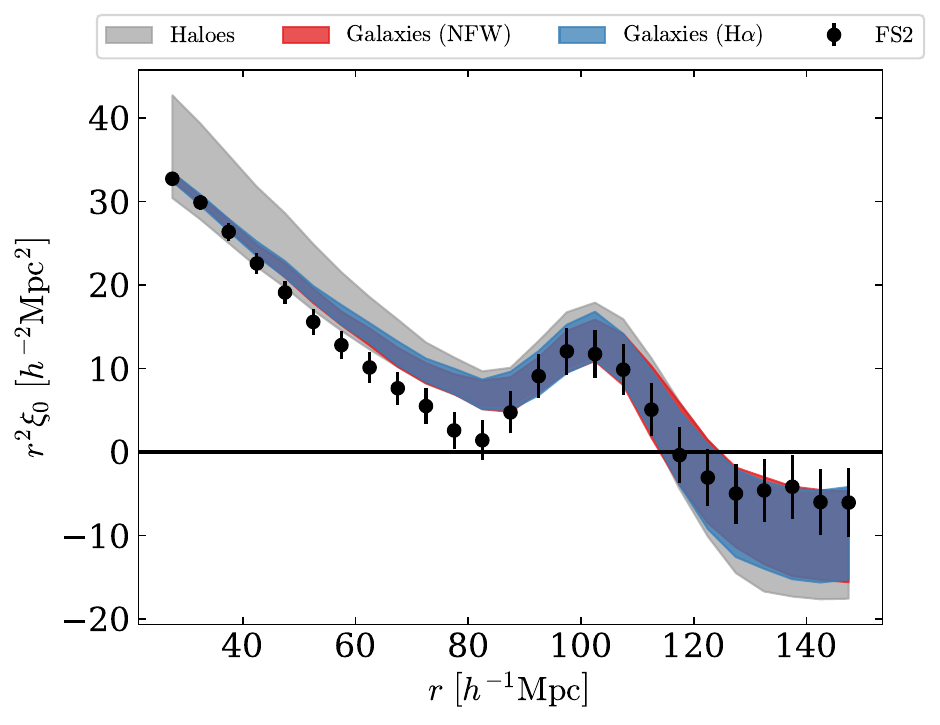}
\caption{Real-space 2PCF of the FS2 spectroscopic sample at $z\in[0.9, 1.1]$ (black points) and the haloes (grey), galaxies with NFW (red) and H$\alpha$ (blue) profiles. The shaded area shows the mean and variance of the signal for every simulation in Table~\ref{tab:bestfit_HOD}.}
\label{fig:realspace2PCF_galaxies}
\end{figure}

Since we analyse snapshots, we compute the real-space 2PCF using the natural estimator \citep{peebles1974statistical}
\begin{equation}
    \xi(s,\mu) = \frac{\NDD(s,\mu)}{\NRR(s,\mu)} - 1\,,
    \label{eq:2PCF_PH_estimator}
\end{equation}
where $\NDD(s,\mu)$ and $\NRR(s,\mu)$ are the normalised counts of data and random pairs in bins of $s$ and $\mu$. We compute \NDD$(s,\mu)$ with \corrfunc \citep{sinha2020corrfunc} while we estimate \NRR($s,\mu$) analytically. To find the HOD parameters to produce the final galaxy catalogue, we run a $\chi^2$-minimisation method,\footnote{As implemented in \texttt{scipy.optimize.differential\_evolution}} with
\begin{equation}
    \chi^2 = \left(\bm{\xi} - \bm{\xi}_{\rm FS2}\right)^{\rm T} \bm{C}_{\rm FS2}^{-1} \left(\bm{\xi} - \bm{\xi}_{\rm FS2}\right)\,,
    \label{eq:chi2}
\end{equation}
where $\bm{\xi}_{\rm FS2}$ and $\bm{C}_{\rm FS2}$ refer to the FS2 lightcone real-space correlation function and covariance on the spectroscopic target H$\alpha$ sample in the redshift bin $z \in [0.9, 1.1]$. 
In practice, we select H$\alpha$ galaxies in the FS2 catalogue by setting a detection threshold of the H$\alpha$ flux at $2\times10^{-16}$\,erg\,s$^{-1}$\,cm$^{-2}$, and an $H$-band magnitude $H < 24$. This generally corresponds to masses above roughly $10^{11}\,h^{-1}\si{\solarmass}$.
To calibrate our HOD model, we impose the final galaxy catalogue to match the mean number density and real-space clustering of the FS2 H$\alpha$ sample for scales $25\mpcoh < s < 150 \mpcoh$. Since the input simulations have different cosmologies than FS2, we prioritised fitting the clustering signal, even at the cost of discrepancies in galaxy bias and satellite fractions.
Last, we note that the mass definition (\Mvirm) of our input halo catalogues is not the same as that in FS2 (\mvir). While it could have been problematic had we needed to perform abundance-matching techniques, our current method only focuses on the clustering properties, in which mass rescaling would be absorbed in the best-fit parameters of our HOD model.
The HOD priors can be found in Table \ref{tab:priors_HOD}, and the best-fit parameters for each simulation in the Appendix~\ref{appendix:hod_parameters} (see Table \ref{tab:bestfit_HOD}).
\begin{table}[]
\caption{Free parameters of our HOD model and their respective priors.}
\smallskip
\label{tab:priors_HOD}
\smallskip
\begin{tabular}{|c|c|}
\hline
\rowcolor{blue!5}
  &  \\[-8pt]
\rowcolor{blue!5}
Parameter & Priors 
\\ \hline
 &  \\[-8pt]
 $\logten(A_{\rm sat,0})$ & $\mathcal{U}~[-4, \logten(2)]$ \\
  $A_{\rm sat,1}$ & $\mathcal{U}~[0, 2]$ \\
  $\mathcal{M}$ & $\mathcal{U}~[10, 15]$ \\
  $\sigma$ & $\mathcal{U}~[0, 10]$ \\
  $\gamma$ & $\mathcal{U}~[-3, 0.5]$ \\
  $\logten\left[M_0/(h^{-1}\si{\solarmass})\right]$ & $\mathcal{U}~[10.5, 15]$ \\ [3pt]
  $\logten\left[M_1/(h^{-1}\si{\solarmass})\right]$ & $\mathcal{U}~[11, 15]$ \\ 
  $\alpha$ & $\mathcal{U}~[0.5, 2]$\\
  \hline
\end{tabular}
\tablefoot{The first column shows the parameters of our HOD model, defined in Eq.~\eqref{eq:ncen} and Eq.~\eqref{eq:nsat}. In the second column we show the uniform priors considered in our $\chi^2$-minimisation procedure to produce galaxy catalogues. $\mathcal{U}[\rm min, max]$ refers to a uniform distribution between min and max values.}
\end{table}
Finally, the mean and variance of the real-space 2PCF across the different simulations are shown in Fig.~\ref{fig:realspace2PCF_galaxies}.
We see that galaxy catalogues (in red or blue) match overall the FS2 clustering, compared to haloes (in grey), as intended. This is particularly visible at lower scales, where the scatter (variance) is much larger for haloes. This shows that our HOD assignment methods work well. We remark, however, that at scales of $r \sim 80\mpcoh$, there is a mismatch between the FS2 signal and the galaxy mocks we produced. As seen in \cite{EuclidSkyFlagship}, this feature very likely comes from an artifact from the FS2 simulation at this redshift, as it cannot be fitted by the theory. This is not a problem in our case, because the HOD parameters are mainly driven by the smallest scales (here $r = 25$\mpcoh), which are accurately recovered.

On larger scales, the halo and galaxy mocks behave similarly, although with smaller variance for galaxies. At these scales, we become dominated by the shot noise, which cannot be completely reduced by an HOD method.
Overall, this shows that our model-independent galaxy-assignment algorithm allows us to match the FS2 clustering signal in real space for H$\alpha$ emitters, thus paving the way for a fair comparison of redshift-space analysis in the following section.

\subsection{Galaxy clustering}
\label{sec:galaxy_clustering}
To compute the galaxy distribution in redshift-space, we displace galaxies along a direction according to their peculiar velocities, snapshot redshift, and background cosmology (given by the simulation) as 
\begin{equation}
    \bm{s} = \bm{x} + \frac{v_{A}(\bm{x})}{a H}\bm{e}_{A}\,,
\end{equation}
where $\bm{s}$ is the redshift-space position, $\bm{x}$ its real-space counterpart, $H$ the Hubble parameter at the snapshot redshift, $v_{A}(\bm{x}) = \bm{v}(\bm{x})\cdot \bm{e}_{A}$ the velocity along $\bm{e}_{A}$, and $\bm{e}_{A} \in \{\bm{e}_{x}, \bm{e}_{y}, \bm{e}_{z}\}$ the unit vector along the directions $x, y$ or $z$. The 2PCF multipoles are estimated as
\begin{equation}
    \xi_\ell(s) = \frac{(2\ell + 1)}{2}\int_{-1}^1 \xi(s, \mu)\,\mathcal{L}_\ell(\mu)\,\diff\mu\,,
\end{equation}
where $\mathcal{L}_\ell$ is the Legendre polynomial of order $\ell$ (which, in our case are $\ell = 0, 2$ and 4), and $\mu = \bm{r}\cdot\bm{e}_{A}$, with $\bm{r}$ the separation for a given galaxy pair. The anisotropic correlation function $\xi(s,\mu)$ is computed with Eq.\,\eqref{eq:2PCF_PH_estimator}, and using the same scales as for the HOD calibration for the three multipoles.
For each galaxy catalogue, we produce an associated Gaussian covariance matrix for the redshift-space 2PCF multipoles. For an explicit expression, see \cite{grieb2016gaussian}. This model needs an input power spectrum that we infer by fitting the \vdg model to the 2PCF multipoles averaged over the three directions of the box. More precisely, we start from a naive covariance matrix and refine its computation over many iterations of the fitting procedure, replacing the covariance with its most recent estimate.

\subsection{Likelihood analysis}

We aim to estimate the best-fit parameters of the \vdg galaxy clustering model (see Sect.~\ref{sec:vdg_theory}) when fitting the 2PCF multipoles of our galaxy catalogues using the template-fitting method. In this case, we fit the data to a `template' by considering a linear power spectrum with a fixed fiducial cosmology that we usually assume to be the real cosmology of our Universe. In this case, we do not vary directly the cosmological parameters, but rather compressed quantities such as $f\sigma_{12}$, the linear growth rate multiplied by the matter density within spheres of radius $12\,$Mpc (which is similar to $f\sigma_8$, but has the advantage of not relying on the value of $h$, which is a desirable property when varying cosmological parameters, $h$ included, \citealt{sanchez2020arguments}), $\alpha_\parallel$, and $\alpha_\perp$, the AP parameters which correspond to distance dilutions along and transverse to the line-of-sight due to the discrepancy between the fiducial cosmology and that of the data.
To do so, we use the nested sampler \multinest \citep{buchner2014multinest} and a Gaussian likelihood similar to Eq.~\eqref{eq:chi2} in redshift space. The parameters and priors are given in Table \ref{tab:priors}.
\begin{table}[]
\caption{Priors on the parameters used in the template-fitting analysis.}
\smallskip
\label{tab:priors}
\smallskip
\begin{tabular}{|c|c|}
\hline
\rowcolor{blue!5}
   &  \\[-9pt]
\rowcolor{blue!5}
 Parameter & Priors \\
\hline
\rowcolor{blue!5}
\multicolumn{2}{|c|}{\bf Template analysis} \\
\hline \\[-8pt]
   $f$ & $\mathcal{U}~[0.5,1.05]$ \\
   $\sigma_{12}$ & $ \mathcal{U}~[0.2,1]$ \\
   $\alpha_\parallel$ & $\mathcal{U}~[0.8,1.2]$ \\
   $\alpha_\perp$ & $\mathcal{U}~[0.8,1.2]$ \\
\hline
\rowcolor{blue!5}
\multicolumn{2}{|c|}{\bf Bias parameters} \\
\hline \\[-8pt]
   $b_1$ & $\mathcal{U}~[0.5,3.5]$ \\
   $b_2$ & $ \mathcal{U}~[-10,10]$ \\
   $b_{\mathcal{G}_2}$ & $\mathcal{U}~[-20,20]$  or fixed to Eq.~\eqref{eq:LL_approximation}\\
   $b_{\Gamma_3}$ & $\mathcal{U}~[-20,20]$  or fixed to Eq.~\eqref{eq:LL_approximation}\\
\hline
\rowcolor{blue!5}
\multicolumn{2}{|c|}{\bf Counterterms} \\
\hline \\[-8pt]
    $c_0\,\left[\left(\si{\hMpc}\right)^2\right]$ & $\mathcal{U}~[-500,500]$ \\[6pt]
    $c_2\,\left[\left(\si{\hMpc}\right)^2\right]$ & $\mathcal{U}~[-500,500]$ \\[6pt]
    $c_4\,\left[\left(\si{\hMpc}\right)^2\right]$ & $\mathcal{U}~[-500,500]$ \\[6pt]
  \hline
\rowcolor{blue!5}
\multicolumn{2}{|c|}{\bf FoG} \\
\hline \\[-8pt]
    $a_{\rm vir}\,[({\rm Mpc}/h)^2]$ & $\mathcal{U}~[0,10]$ \\
  \hline
\end{tabular}
\tablefoot{The AP parameters are \citep{alcock1979evolution}
\begin{equation}
    \alpha_{\parallel}  = \frac{H'(z)}{H(z)}\,, \quad {\rm and} \quad \alpha_{\perp}  = \frac{d_A(z)}{d_A'(z)} \, ,
    \label{eq:AP_parameters}
\end{equation}
where a prime denotes an estimation at the fiducial cosmology, $H$ is the Hubble parameter and $d_A$ the angular diameter distance. Regarding the bias parameters, $b_1$ and $b_2$ are the linear and quadratic bias, while $b_{\mathcal{G}_2}$ and $b_{\Gamma_3}$ are non-local biases. The counterterms are given by $c_0$, $c_2$, and $c_4$, while Finger-of-God effects are described by $a_{\rm vir}$.}
\end{table}

We also test the robustness of our modelling under the local-Lagrangian approximation, using co-evolution relations for the higher-order bias (\citealt{karcher20262PCF}, and references therein)
\begin{equation}
    b_{\mathcal{G}_2} = -\frac{2}{7}b_1\,, \quad 
    ~b_{\Gamma_3} = \frac{11}{42}b_1\,, \quad 
    ~b_{s^2} = -\frac{4}{7}b_1\,,
\label{eq:LL_approximation}
\end{equation}
with $b_1$ the linear Lagrangian bias, related to the linear Eulerian bias through $b_1 = 1 + b_1^{\rm L}$.

We then extract the best-fit model and 68\% confidence interval for the template-fitting parameters,  marginalising over the nuisance parameters  using \getdist \citep{lewis2019getdist}.
The recovered parameters are then compared to their fiducial values. We note that the AP parameters should be equal to unity when the fiducial cosmology is the same as that of the data. 
For most models, we use \comet~ to estimate the fiducial values of the linear growth rate and AP parameters. 
Some models, such as $f(R)$ and interacting dark energy models, are not supported by \comet and feature scale-dependent growth rates, meaning that, strictly speaking, template fitting cannot be applied. 
In this case, we use an approximated growth rate, estimating $f = \partial\ln D_+/\partial\ln a$, with $D_+$ the growth factor, using the matter power spectrum at different redshifts and averaging between the scales $k_{\rm min} = 2\pi/L_{\rm box}$ and $k_{\rm max} = 0.1\homopc$.
Massive neutrinos also lead to a scale-dependent growth rate, but, in this case, we directly use \comet, which computes the large-scale limit and where the massive neutrino contribution appears in the total matter energy budget. Applying template-fitting methods with theoretical models that account for a scale-dependent growth rate is left for future work.

\section{\label{sec:Results}Results}
In this section, we present the results of our template-fitting analysis of our galaxy catalogues in snapshots a $z = 1$, first using a template with the same cosmology as the data, then using an FS2 template (see Sect.~\ref{sec:fs2}). For each simulation, we consider the axis for which the redshift-space 2PCF multipoles are closest to the average. This is because different snapshot axes can lead to different recoveries of cosmological parameters, sometimes very far from their expected values. In practice, averaging the multipoles along the three directions significantly reduces the variance, allowing us to recover an unbiased estimate \citep{smith2021reducing}. However, the averaging procedure also has a non-trivial impact on the covariance, which is not simply reduced by a third. One possibility was to average the 2PCF multipoles and compute the associated covariance matrix. We chose instead to select, for each case, the axis which is the closest to the average (which gives the smallest $\chi^2$), allowing us to keep the covariance matrix described in Sect.~\ref{sec:galaxy_clustering}. For completeness, we also show the results in all directions in Appendix~\ref{appendix:all_directions}.

\subsection{2PCF template fitting in redshift space}

\begin{figure*}[htbp!]
\centering
\includegraphics[width=1.0\hsize]{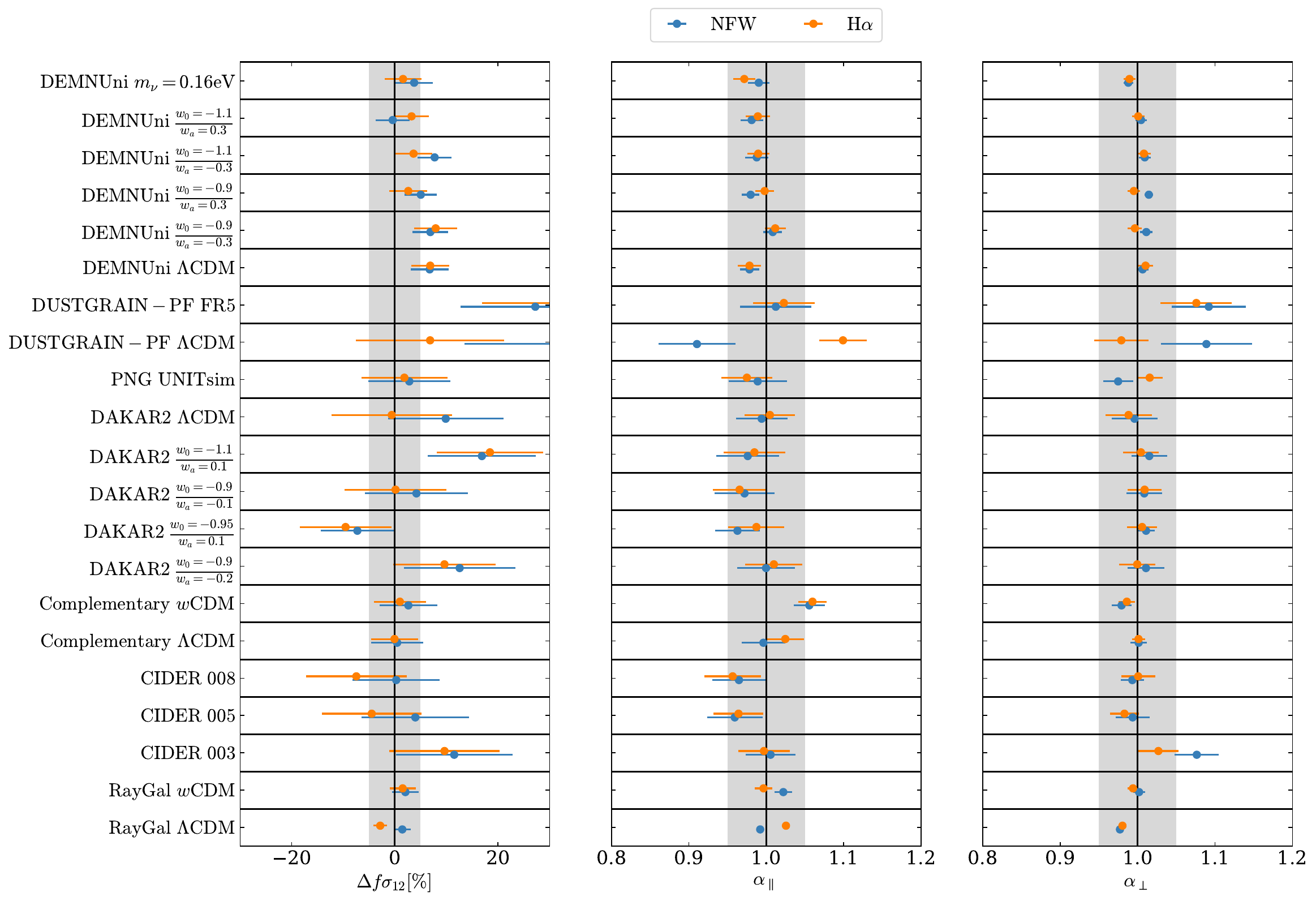}
\caption{Marginalised constraints on the template-fitting parameters of the \vdg model with NFW (blue) and H$\alpha$ (orange) profiles, using the same cosmology for the simulation and the template. We show the relative difference $\Delta f\sigma_{12}$ on the $f\sigma_{12}$ parameter with respect to its fiducial value, and the absolute values of the dilation parameters $\alpha_\perp$ and $\alpha_\parallel$. The grey shaded area shows the $\pm$ 5\% limits.}
\label{fig:results_template_same_LL_bestaxis}
\end{figure*}

We first consider the case where the fiducial cosmology in the template-fitting analysis is that of the simulation. The results can be found in Fig.~\ref{fig:results_template_same_LL_bestaxis}.
We see that overall, the AP parameters are well recovered to a value close to unity. For $f\sigma_{12}$, we also find an agreement, although with larger scatter. Because all the catalogues have the same mean galaxy density, the size of the error bars mainly comes from the simulation volumes (as well as possible parameter degeneracies).
It seems that the \DUSTGRAINPATHFINDER simulations do not recover the fiducial parameters well. It can be noted that these are the smallest simulations in terms of box size and number of particles, thus explaining the larger error bars. This discrepancy could thus be attributed to larger variance, even if the error bars are not compatible. This suggests that the chosen direction is more biased than others, as we can see in Appendix\,\ref{appendix:all_directions}.

Finally, we see that the NFW and H$\alpha$ profiles lead to a similar parameter recovery, except, again, for the \DUSTGRAINPATHFINDER simulations. This indicates that the discrepancies come more from a given realisation of galaxy assignments, which leads to stochastic fluctuations of the galaxy clustering signal. Other than that, it does not seem that there is a clear difference between the density profiles considered, which makes sense given the scales we are interested in.

\subsection{\Euclid Flagship template}

\begin{figure*}[htbp!]
\centering
\includegraphics[width=1.0\hsize]{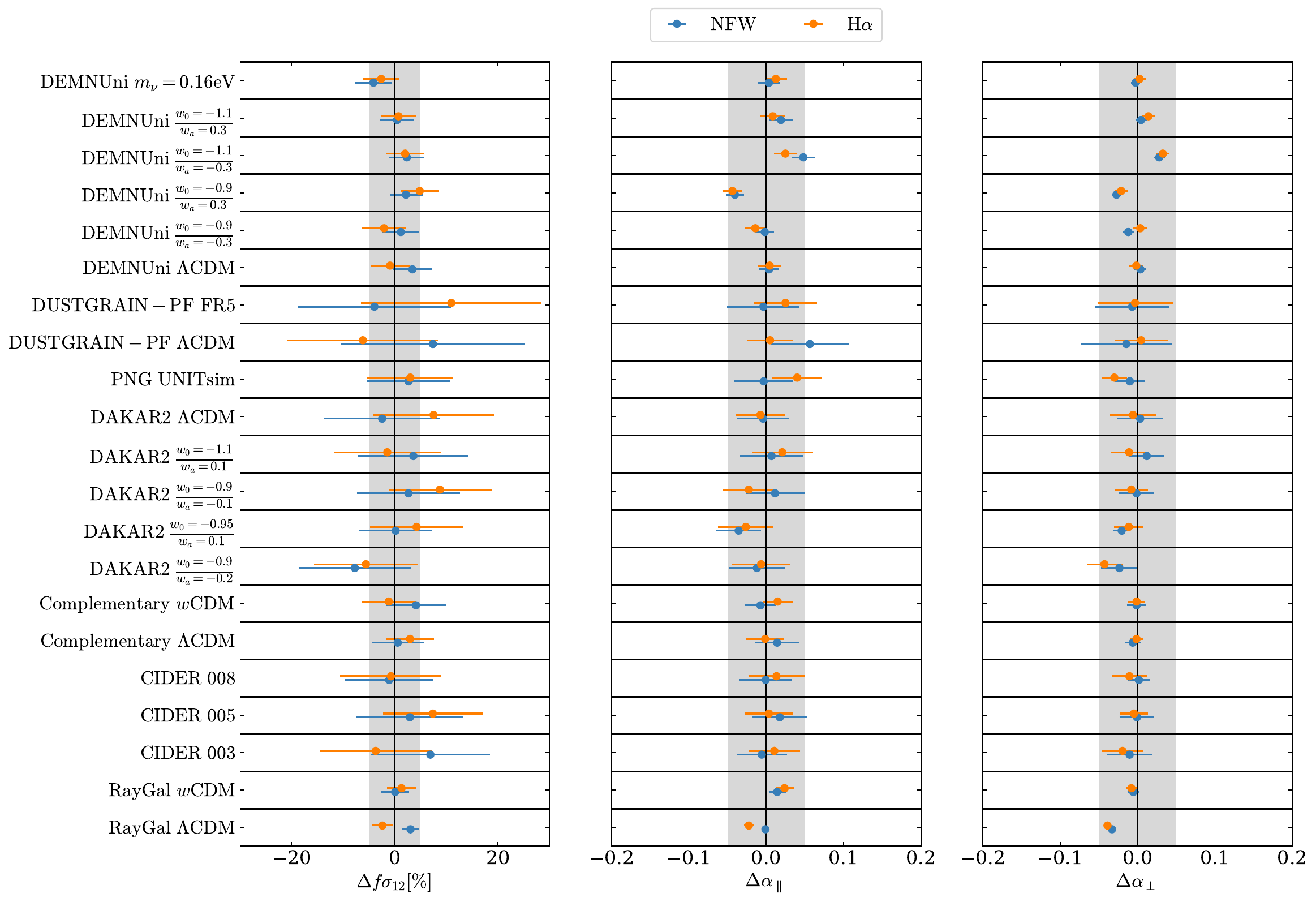}
\caption{Same as Fig.~\ref{fig:results_template_same_LL_bestaxis} but using a template with \FLAGSHIPTWO cosmology. We add a correction equal to the deviation found in Fig.\,\ref{fig:results_template_same_LL_bestaxis}. For the AP parameters, we subtract the fiducial values using Eq.~\eqref{eq:AP_parameters}.}
\label{fig:results_template_fs2_LL_bestaxis}
\end{figure*}

Next, we perform the same analysis, but using this time a fixed template with the FS2 cosmology shown in Table~\ref{tab:flagship_cosmology}. The goal is to assess whether or not we are able to recover the correct cosmological parameters for a wide range of models, which can vastly differ from that of the template. Such models possess different cosmological parameters, as well as different gravity theories. The results are shown in Fig.~\ref{fig:results_template_fs2_LL_bestaxis}.
Because we focus specifically on the impact of the template, we subtract the deviations found in Fig\,\ref{fig:results_template_same_LL_bestaxis}, which should be due to statistical fluctuations and are, in principle, also present when modifying the template. 

First, we remark that in any case, deviations from fiducial values are much smaller than in Fig.\,\ref{fig:results_template_same_LL_bestaxis}, meaning that most of the error comes from statistical fluctuations. For example, we see that the estimated values for $f\sigma_{12}$ agree with their fiducial values within error bars.  
Second, for most simulations, it is difficult to conclude that the best-fit AP parameters are closer to their expectation value, rather than unity, even if the scatter seem smaller. The reason is that even if the exact cosmology of these simulations is different from FS2, they are still quite close overall. This is because, in the last decade, it has been common to use cosmologies directly derived from \cite{planck2016cosmological} or \cite{planck2018cosmological}. Furthermore, for every simulation, we decreased the galaxy density to match that of H$\alpha$ emitters at the redshift of interest. For most simulations, the sample variance appears to be of the same order or larger than the shift in the AP parameters.

Fortunately, this is not the case for the \RAYGAL simulations, whose cosmology is that of WMAP-7yr, and is therefore vastly different from FS2. For the \wcdm \RAYGAL simulation, $f\sigma_{12}$ and $\alpha_\perp$ are very accurately recovered within error bars, while we see a 2\,$\sigma$ deviation on $\alpha_\parallel$. For the \lcdm \RAYGAL simulation, we exactly recover $\alpha_\parallel$, but find a deviation in $\alpha_\perp$ that was already present in Fig\,\ref{fig:results_template_same_LL_bestaxis}, but with larger amplitude. However, due to the data size or resolution, it is difficult to conclude whether this is an issue of covariance (that we produced using the Gaussian approximation), or a hint towards a failure of the template-fitting method. 

Regarding $f\sigma_{12}$, the NFW case, which was slightly larger than the expected value when using the correct cosmology for the template, becomes even larger. The opposite occurs for the H$\alpha$ case, where the value was smaller than expected and becomes even smaller with the FS2 template. In both cases, we see that if we were averaging over the two profiles, we would recover an unbiased estimate for $f\sigma_{12}$ for both the FS2 template and for the simulation template. The lack of a consistent visible trend between NFW and H$\alpha$ profiles are due to the fact that the scales of the present analysis are more sensitive to the 2-halo term, and therefore less to the density profile.
This indicates that fluctuations are mainly due to a given realisation of the catalogues, which can be mistaken for a wrong cosmology, that is, a shift in the AP parameters. As a consequence, it also impacts the estimation of the growth rate, but not by a large amount.

Interestingly, this means that, at least for the range of simulations we considered, the estimation of the growth rate is not very sensitive to the template, even for the case where the cosmology of the data has $\Omega_{\rm m,0} \approx 0.26$ (\lcdm \RAYGAL) and that of the template $\Omega_{\rm m,0} \approx 0.32$ (FS2). Regarding the AP parameters, they can usually be recovered in the best-case scenario, even for vastly different cosmologies (\wcdm \RAYGAL compared to FS2), although one has to be careful about the covariance model.

\section{\label{sec:Conclusion}Conclusions}

We produced galaxy catalogues for numerous \lcdm and non-standard simulations, that match the signal of target H$\alpha$ galaxies of the \Euclid spectroscopic sample. We then performed a galaxy clustering template-fitting analysis in redshift space, assessing the impact of the template on parameter recovery.

As inputs, we used the halo catalogues produced with a modified version of \rockstar in \cite{EP-Racz}. In terms of non-standard simulations, we considered dynamical dark energy models such as \wcdm, \wwcdm, massive neutrinos, primordial non-Gaussianity, interacting dark energy, and the \cite{hu2007models} $f(R)$-gravity model. We assigned galaxies to haloes using the HOD framework and considered a very flexible HOD model with conformity relations for the satellite distribution. For completeness, we considered two possibilities for halo density profiles: NFW and the H$\alpha$ fitted by \cite{reyes-peraza2024improved}. In both cases, we accounted for the triaxial halo shapes. We constrained the HOD parameters by imposing that the real-space 2PCF of the output galaxy catalogues must be similar to that of the FS2 H$\alpha$ galaxy catalogue on the lightcone on the scales of interest for galaxy clustering analyses. That is to ensure that the intrinsic clustering on large scales is similar to that we expect from \Euclid data, without constraining too much the small scales and velocities, which are in general more sensitive to non-standard models. Finally, we also imposed that the mean galaxy density matches that of H$\alpha$ galaxies, as expected by the model 3 of \cite{pozzetti2016modelling}.
The advantage of these catalogues is that we processed a wide range of simulations and models through the same model-agnostic pipeline, and ensured that it matches observations. These catalogues are made publicly available through CosmoHub.

Beyond the catalogue data release, the science goal of this paper was to investigate the limits of the template-fitting approach for galaxy clustering. We fitted the multipoles of the 2PCF in redshift space
to estimate $f\sigma_{12}$ as well as the AP parameters $\alpha_\parallel$ and $\alpha_\perp$, along with bias and nuisance parameters. For the modelling, we used the \vdg model \citep{sanchez2017clustering}, and its implementation in the \comet emulator \citep{eggemeier2023comet, eggemeier2025boosting}, taking advantage of the fact that it implements massive neutrinos \citep{pezzotta2025extending}. This model has been shown to be one of the best in the model-comparison challenge performed in \citet{karcher20262PCF}.
While this was already known to some extent, template-fitting methods are sometimes referred to as `model-independent', which is not exactly true because an extremely wrong template can lead to biased parameter recovery. The question is rather: how wrong can we afford to be in the template, so that it does not lead to a large bias? This is an important question, as we do not know the real cosmology of our Universe. We thus perform the galaxy-clustering analysis using first a template which has the same cosmology as the data: this is a null test to assess the level of parameter recovery we can reach; and second, an FS2 template which is close to \cite{planck2018cosmological}, to estimate the impact of the difference in cosmology. We found that overall, we had a parameter recovery better than 5\% when the cosmology template was the same as the data, except for the \DUSTGRAINPATHFINDER simulations which seem to overestimate $f\sigma_{12}$. Furthermore, although there are differences between the NFW and H$\alpha$ profiles, there is no consistent trend, and thus this discrepancy mainly comes from sample variance. When using the FS2 template, we did not find any large difference in the $f\sigma_{12}$ compared to before, showing that this parameter is very robust to any change in the template. For the AP parameters, it seems that in general it is close to the expectation value, although some deviations much larger than error bars can be found, meaning that one has to be careful about covariance modelling. The AP parameters estimated from 2PCF multipoles should thus be taken with care and cross-checked with more refined BAO analysis.

There are several limitations to the present work, which should be accounted for in future studies in order to further confirm our conclusions. The first limitation comes from the set of simulations we consider: except for the \RAYGAL simulations, all of them are already very close to FS2 in terms of $\Omega_{\rm m,0}$, which is the main parameter that drives the discrepancy in terms of the growth rate or AP parameters. While it is understandable that scientists prefer to run simulations close to the most accurate, state-of-the-art model, it would certainly have value to also run simulations with more exotic models in terms of gravity and dark energy, as well as significantly different cosmological parameters -- especially $\Omegam$. This would allow us to address potential severe issues in our analysis, such as the one seen in the AP parameter recovery. 
Another limitation comes from both the relatively low galaxy density in the target sample and the overall low resolution of simulations or small volumes. The latter comes from the fact that non-standard simulations are generally more expensive than \lcdm, and cannot reach the same resolution. In our case, most simulations have a particle mass $m_{\rm p} \geq 10^{10} \msoh$, which is very large, and using $m_{\rm p} \leq 10^{9} \msoh$ would be more desirable to better characterise H$\alpha$ emitters. Often, the 2PCF of our halo catalogues already has a larger amplitude than the target galaxy 2PCF, meaning that our HOD model must sometime largely decrease the number of objects. The low density of the target sample further forces us to remove galaxies from our catalogues, leading to large shot noise. It is important, for the future, to analyse high-resolution simulations with large volumes (with box lengths of at least $2\,\gpcoh$) in a wide range of models in order to make our findings insensitive to sample variance.

In this paper, we only considered simulation snapshots at $z = 1$. An extension of the present work would include exploring the redshift dependence and its relation with the galaxy clustering modelling. Furthermore, because we only had access to simulation snapshots, we focused the present analysis on galaxy clustering. Opening to the lightcone would necessitate calibrating our HOD and analysing our data not only on the clustering signal, but on the full 3$\times$2 point statistics at least. This would generalise the methodology described in this paper to be the closest to observations, and solidify our trust in our models.

\section{Data availability}
\label{sec:data_availability}

The galaxy catalogues are available on the Cosmohub platform, under the name `Non-standard cosmological model'.

\begin{acknowledgements}
MAB acknowledges support from project “Advanced Technologies for the exploration of the Universe”, part of Complementary Plan ASTROHEP, funded by the European Union - Next Generation (MCIU/PRTR-C17.I1).
SA has been funded by MCIN/AEI/10.13039/501100011033 and FSE+ (Europe) under project PID2024-156844NA-C22 and the RYC2022-037311-I fellowship.
This project was provided with computer and storage resources by GENCI at TGCC thanks to the grant 2023-A0150402287 on Joliot Curie's SKL partition.
This work has made use of CosmoHub. CosmoHub has been developed by the Port d'Informació Científica (PIC), maintained through a collaboration of the Institut de Física d'Altes Energies (IFAE) and the Centro de Investigaciones Energéticas, Medioambientales y Tecnológicas (CIEMAT) and the Institute of Space Sciences (CSIC \& IEEC).
CosmoHub was partially funded by the "Plan Estatal de Investigación Científica y Técnica y de Innovación" program of the Spanish government, has been supported by the call for grants for Scientific and Technical Equipment 2021 of the State Program for Knowledge Generation and Scientific and Technological Strengthening of the R+D+i System, financed by MCIN/AEI/ 10.13039/501100011033 and the EU NextGeneration/PRTR (Hadoop Cluster for the comprehensive management of massive scientific data, reference EQC2021-007479-P) and by MICIIN with funding from European Union NextGenerationEU(PRTR-C17.I1) and by Generalitat de Catalunya.
GR acknowledges the support of the Research Council of Finland grant 354905 and the support by the European Research Council via ERC Consolidator grant KETJU (no. 818930).

\AckEC  

\end{acknowledgements}

%
%

\bibliography{my, Euclid}

%

\begin{appendix}

\section{Halo occupation distribution parameters}
\label{appendix:hod_parameters}
In this section, we present the best-fit parameters of the HOD model described in Sect.~\ref{sec:galaxy_catalogue_production}, for all the simulations in Table~\ref{tab:sims}. The results are shown in Table~\ref{tab:bestfit_HOD}.
\begin{table*}[]
    \centering
     \caption{Galaxy catalogue characteristics for each simulation snapshot at $z = 1$ and satellite profile considered: the number density ratio with respect to the target value for H$\alpha$ emitters \citep{pozzetti2016modelling} and best-fit parameters of our HOD model.}
    \begin{tabular}{|c c | c c c c c c c c|}
  \hline
  \rowcolor{blue!5}
  & & & & & & & & &\\[-8pt]
  \rowcolor{blue!5}
 Name & Satellite profile & $A_{\rm sat,0}$ & $A_{\rm sat,1}$ & $\mathcal{M}$  & $\sigma$ & $\gamma$ & $\logten\frac{M_0}{h^{-1}\si{\solarmass}}$ &  $\logten\frac{M_1}{h^{-1}\si{\solarmass}}$ & $\alpha$ 
 \\ \hline
    & & & & & & & & &\\[-8pt]
    \multirow{2}{*}{\RAYGAL \lcdm}  & NFW & $4.66\times10^{-4}$ & $1.99$ & 11.5 & 7.98 & $-2.22$& 10.6 & 11.2 & 0.57 \\
      & H$\alpha$& $1.58\times10^{-4}$ & $1.76$ & 11.4& 8.37& $-2.44$ & 10.6 & 11.2 & 0.96 
    \\ \hline
    & & & & & & & & &\\[-8pt]
    \multirow{2}{*}{\RAYGAL \wcdm}  & NFW & $8.94\times10^{-4}$ & 0.81 & 12.2 & 6.39 & $-2.73$ & 10.8 & 11.8 & 1.21 \\
      & H$\alpha$& $1.27\times10^{-4}$& 1.12 & 12.2 & 9.17& $-2.18$ & 11.4 & 11.6 & 0.82
      \\ \hline
      & & & & & & & & &\\[-8pt]
    \multirow{2}{*}{\COMPLEMENTARY \lcdm}  & NFW & $2.87\times10^{-3}$ & 1.56 & 14.8 & 3.76 & $-1.59$ & 12.8 & 13.8 & 1.01 \\
      & H$\alpha$& $3.51\times10^{-4}$ & 0.99 & 13.4 & 8.27 & $-1.61$ & 11.2 & 13.1 & 1.46 
      \\ \hline
      & & & & & & & & &\\[-8pt]
    \multirow{2}{*}{\COMPLEMENTARY \wcdm}  & NFW & $7.72\times10^{-3}$ & 0.46 & 14.9 & 5.29 & $-0.88$ & 12.0 & 14.0 & 1.22\\
      & H$\alpha$& $3.11\times10^{-2}$& 0.64 & 14.4 & 4.90 & $-0.87$ & 11.5 & 14.0 & 1.10 
      \\ \hline 
      & & & & & & & & &\\[-8pt]
    \multirow{2}{*}{\PNGUNITsim}  & NFW & $1.45\times10^{-4}$ & 0.99 & 15.0 & 1.55 & $-2.12$ & 13.5 & 14.4 & 1.83 \\
      & H$\alpha$& $2.62\times10^{-3}$ & 0.50 & 14.5 & 1.44 & $-2.22$ & 11.5 & 14.2 & 0.95 
      \\ \hline
      & & & & & & & & &\\[-8pt]
    \multirow{2}{*}{\DUSTGRAINPATHFINDER \lcdm}  & NFW & $2.74\times 10^{-3}$ & 0.16 &  12.9 &  1.00 &
 $-2.39$ &  11.0 &  12.6 &  0.87 \\
      & H$\alpha$& $6.24\times 10^{-3}$ &  0.06 &  13.0 &  4.81 &
 $-1.72$ &  11.9 &  12.7 &  0.73 
 \\ \hline 
 & & & & & & & & &\\[-8pt]
    \multirow{2}{*}{\DUSTGRAINPATHFINDER FR5}  & NFW & $1.24\times 10^{-4}$  & 0.01 &  13.2 &  5.45 &
 $-1.32$ &  11.1 &  12.5 &  0.77 \\
      & H$\alpha$& $2.34\times 10^{-3}$  & 0.15 &  13.3 &  9.83 &
 $-1.40$ &  12.2 &  13.1 &  1.37 
 \\ \hline
 & & & & & & & & &\\[-8pt]
    \multirow{2}{*}{\CIDER/\DAKARTWO \lcdm}  & NFW & $4.43\times10^{-3}$ & 0.29 & 12.7 & 3.42 & $-2.70$ & 10.8 & 12.7 & 0.97 \\
      & H$\alpha$& $1.91\times10^{-3}$ & 0.14 & 12.5 & 5.44 & $-1.62$ & 12.1 & 12.5 & 0.80 
      \\ \hline 
      & & & & & & & & &\\[-8pt]
    \multirow{2}{*}{\CIDER 003}  & NFW & $8.31\times10^{-4}$ & 0.05 & 12.5 & 4.99 & $-2.23$ & 11.7 & 12.1 & 0.65 \\
      & H$\alpha$& $1.40\times10^{-3}$ & 0.09 & 12.6 & 5.68 & $-2.56$ & 12.0 & 12.5 & 1.21 
      \\ \hline
      & & & & & & & & &\\[-8pt]
    \multirow{2}{*}{\CIDER 005}  & NFW & $2.72\times10^{-2}$ & 0.75 & 12.3 & 9.52 & $-2.96$ & 11.2 & 12.3 & 0.80 \\
      & H$\alpha$ & $1.17\times10^{-4}$ & 0.17 & 12.4 & 4.40 & $-1.96$ & 11.7 & 12.3 & 0.64 
      \\ \hline
      & & & & & & & & &\\[-8pt]
    \multirow{2}{*}{\CIDER 008}  & NFW & $6.97\times10^{-4}$ & 0.62 & 12.0 & 4.97 & $-2.71$ & 11.8 & 11.8 & 0.57 \\
      & H$\alpha$& $1.54\times10^{-4}$ & 0.20 & 12.0 & 1.71 & $-2.64$ & 11.7 & 11.8 & 0.86 
      \\ \hline
      & & & & & & & & &\\[-8pt]
    \multirow{2}{*}{\DAKARTWO $\frac{w_0 = -0.9}{w_a = -0.2}$}  & NFW & $3.27\times10^{-3}$ & 0.05 & 12.7 & 6.61 & $-1.83$ & 11.1 & 12.7 & 1.56 \\
      & H$\alpha$& $9.52\times10^{-3}$ & 0.65 & 12.5 & 6.80 & $-2.32$ & 12.2 & 12.4 & 0.82 
      \\ \hline 
      & & & & & & & & &\\[-8pt]
    \multirow{2}{*}{\DAKARTWO $\frac{w_0 = -0.95}{w_a = 0.1}$}  & NFW & $2.48\times10^{-4}$ & 0.19 & 12.2 & 7.77 & $-1.90$ & 11.8 & 12.1 & 1.09 \\
      & H$\alpha$& $1.05\times10^{-4}$ & 0.03 & 12.4 & 3.28 & $-1.24$ & 10.8 & 11.4 & 0.51 
      \\ \hline
      & & & & & & & & &\\[-8pt]
    \multirow{2}{*}{\DAKARTWO  $\frac{w_0 = -0.9}{w_a = -0.1}$}  & NFW & $2.57\times10^{-3}$ & 0.02 & 12.6 & 2.53 & $-2.22$ & 11.7 & 12.2 & 1.02 \\
      & H$\alpha$& $8.39\times10^{-4}$ & 0.02 & 12.6 & 7.97 & $-1.93$ & 12.1 & 12.1 & 0.86 
      \\ \hline
      & & & & & & & & &\\[-8pt]
    \multirow{2}{*}{\DAKARTWO  $\frac{w_0 = -1.1}{w_a = 0.1}$}  & NFW & $6.28\times10^{-4}$ & 0.13 & 12.6 & 9.59 & $-2.55$ & 12.2 & 12.3 & 1.46 \\
      & H$\alpha$& $3.72\times10^{-2}$ & 0.03 & 12.3 & 6.61 & $-1.12$ & 11.5 & 12.2 & 0.56 
      \\ \hline
      & & & & & & & & &\\[-8pt]
    \multirow{2}{*}{\DEMNUni \lcdm}  & NFW & $1.45\times 10^{-4}$ & 0.45 &  12.8 & 5.39 &
 $-2.53$ &  11.0 &  12.3 &  1.01 \\
      & H$\alpha$& $1.75\times 10^{-3}$ &  0.54 & 12.8 &  4.54 &
 $-3.00$ & 11.7  & 12.2 & 0.52 
    \\ \hline 
    & & & & & & & & &\\[-8pt]
    \multirow{2}{*}{\DEMNUni $\frac{w_0 = -0.9}{w_a = -0.3}$}  & NFW & $3.57\times10^{-4}$ & 0.47 &  12.7 &  1.89 &
 $-1.40$ &  11.3 &  12.7 &  0.95 \\
      & H$\alpha$& $9.37\times 10^{-3}$ & 0.83 &  12.4 &  8.83 &
 $-1.11$ &  10.9 &  12.3 &  0.56 
    \\ \hline
    & & & & & & & & &\\[-8pt]
    \multirow{2}{*}{\DEMNUni $\frac{w_0 = -0.9}{w_a = 0.3}$}  & NFW & $1.68\times 10^{-3}$ & 0.74 &  12.6 & 1.65 &
 $-2.27$ & 11.0  & 12.5 & 1.18 \\
      & H$\alpha$& $6.07\times10^{-3}$ & 0.19 &  12.6 &  6.17 &
 $-1.83$ &  11.3 &  11.5 &  0.56 
    \\ \hline
    & & & & & & & & &\\[-8pt]
    \multirow{2}{*}{\DEMNUni $\frac{w_0 = -1.1}{w_a = -0.3}$}  & NFW & $6.61\times 10^{-2}$ &  1.96 & 12.5 & 7.97 & $-1.55$ & 11.7
 & 12.5 & 0.66 \\
      & H$\alpha$& $7.19\times10^{-3}$  & 0.75 &  12.7 &  9.00 &
 $-2.05$ &  12.0 &  12.5 & 0.75 
    \\ \hline
    & & & & & & & & &\\[-8pt]
    \multirow{2}{*}{\DEMNUni $\frac{w_0 = -1.1}{w_a = 0.3}$}  & NFW & $3.83\times 10^{-2}$ &  0.86 & 12.6 & 4.33 & $-2.19$ & 11.7 &
 12.5 &  0.75 \\
      & H$\alpha$& $1.70\times 10^{-3}$ & 0.17 &  12.9 &  6.42 &
 $-2.08$ &  11.4 &  12.2 &  0.55 
    \\ \hline
    & & & & & & & & &\\[-8pt]
    \multirow{2}{*}{\DEMNUni $m_\nu = 0.16$ eV}  & NFW & $2.43\times 10^{-4}$ &  0.35 &  12.4 &  0.632 &
 $-1.66$ &  11.4 &  11.7 &  0.91 \\
      & H$\alpha$& $3.08\times10^{-4}$ & 0.59 &  12.5 &  7.34 &
 $-2.36$ &  10.6 &  11.5 &  0.54 \\
    \hline
    \end{tabular}
    \label{tab:bestfit_HOD}
\end{table*}
We see that $\mathcal{M}$ is quite low, especially for simulations with large particle mass. In this case, the mean central galaxy distribution peaks early and decays for larger masses. This is due to the fact that the halo 2PCF has a larger bias than our target galaxy catalogue, and our HOD model must decrease the bias by removing the high-mass haloes. This is the reason why we opted for this flexible HOD model.
This also implies very small values of $A_{\rm sat,0}$, ensuring that haloes without any central galaxy do not produce any large number of satellites. Interestingly, there is no apparent specific trend on $A_{\rm s,1}$, meaning that the conformity relation, focussing the satellite distribution in haloes that already contain a central galaxy, is a good selection. Because most of the central galaxies are in haloes of mass $10^\mathcal{M}\,h^{-1}\si{\solarmass}$, so will the satellites, and $\logten\left[M_1/(h^{-1}\si{\solarmass})\right] \approx \mathcal{M}$.

\section{All directions}
\label{appendix:all_directions}

\begin{figure*}[htbp!]
\centering
\includegraphics[width=1.0\hsize]{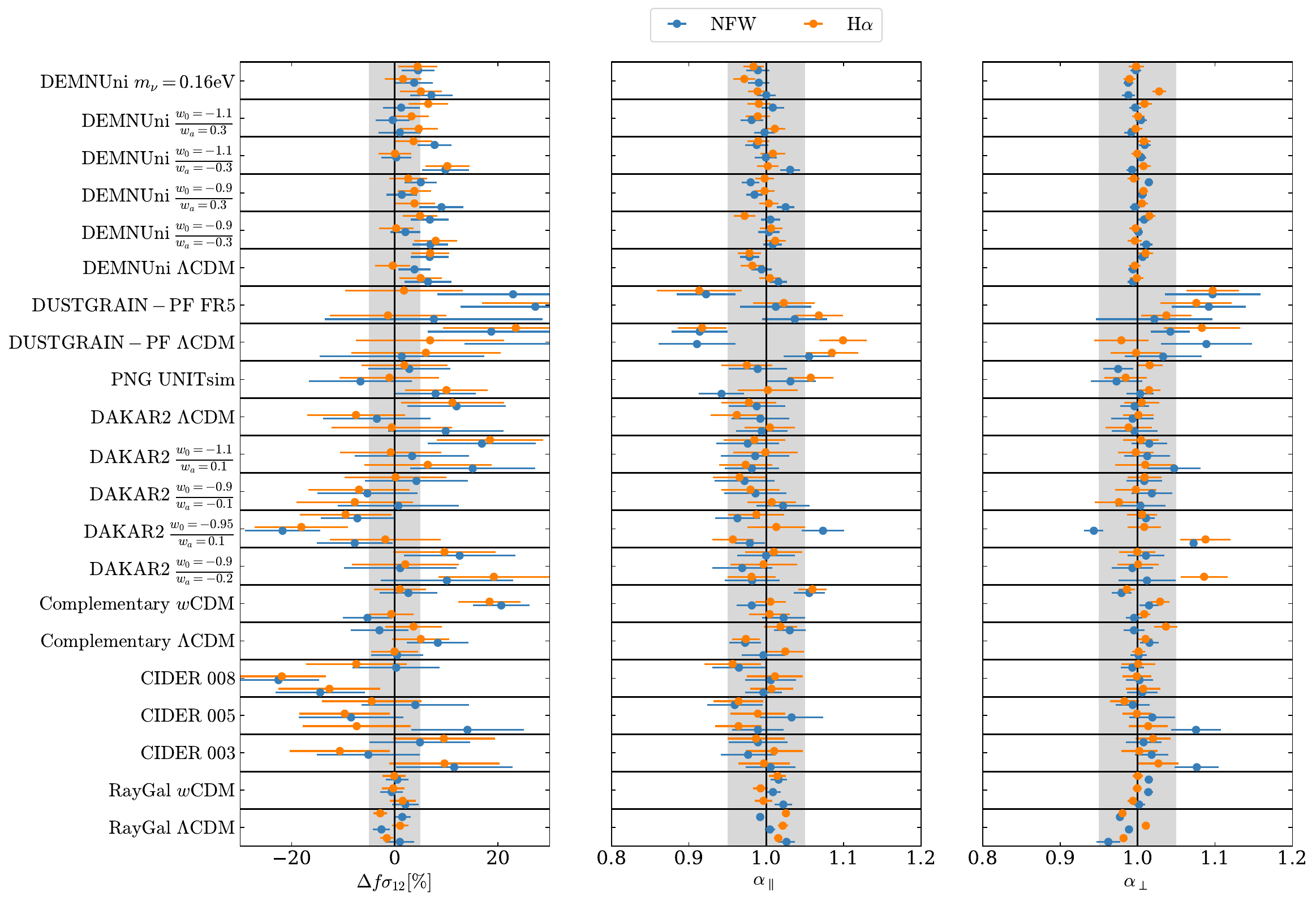}
\caption{Same as Fig.~\ref{fig:results_template_same_LL_bestaxis}, but where each point is duplicated three times, according to the redshift-space distortion projection along the $x$, $y$, and $z$ axes (from bottom to top).}
\label{fig:results_template_same_LL_allaxis}
\end{figure*}

\begin{figure*}[htbp!]
\centering
\includegraphics[width=1.0\hsize]{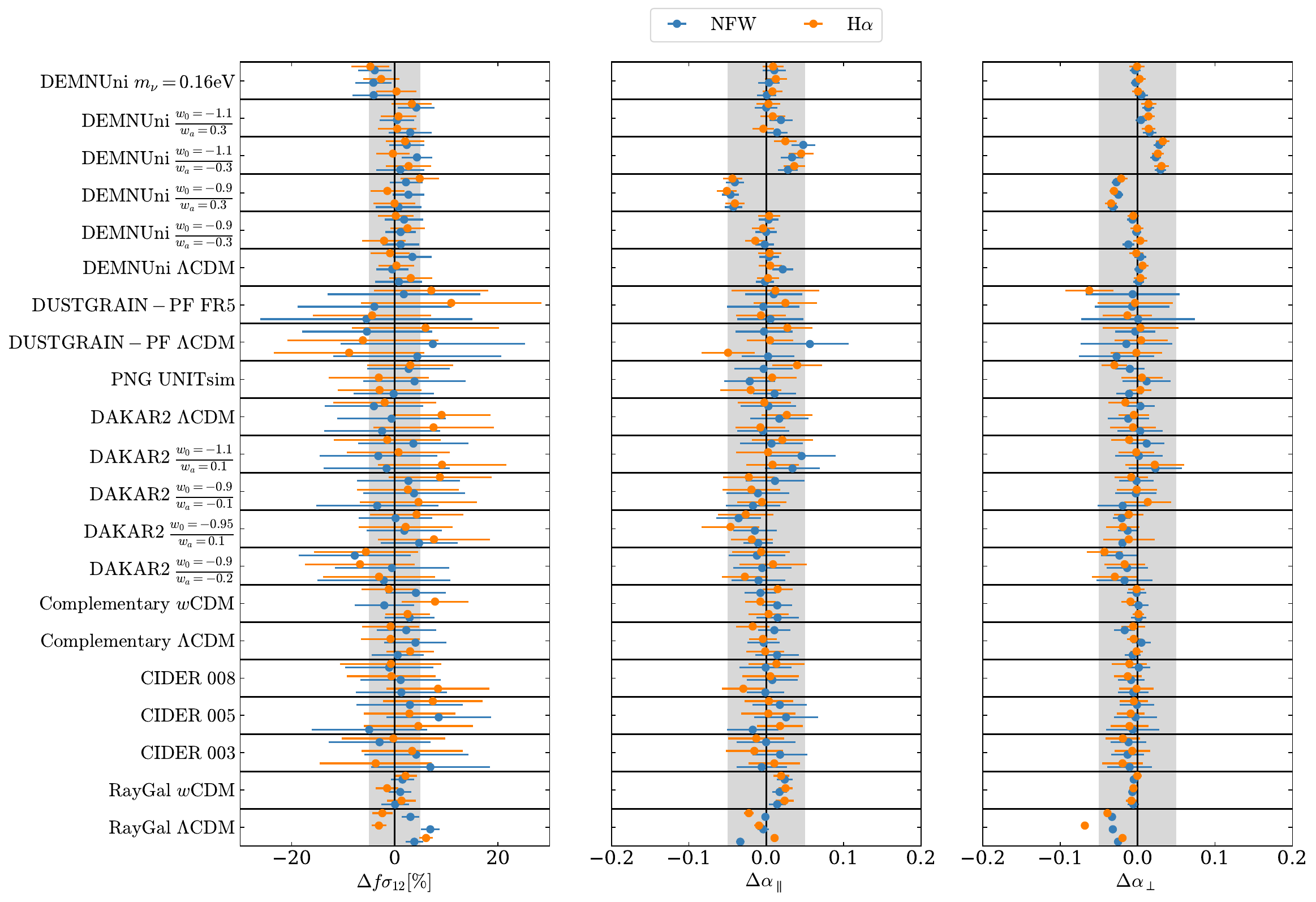}
\caption{Same as Fig.~\ref{fig:results_template_same_LL_allaxis} but using a template with the simulations cosmology.}
\label{fig:results_template_fs2_LL_allaxis}
\end{figure*}

In the main text, in Sect.~\ref{sec:Results}, we only showed the snapshot axis that was closest to the average. Our intent was to minimise the effect of bias due to the line of sight to focus primarily on the impact of the template, and to produce a more readable plot, which is already very busy due to the large number of simulations considered. For completeness, in this appendix, we show the same plots as in Figs.~\ref{fig:results_template_same_LL_bestaxis} and \ref{fig:results_template_fs2_LL_bestaxis}, but this time we are adding the other directions. The results are shown in Figs.~\ref{fig:results_template_same_LL_allaxis} and \ref{fig:results_template_fs2_LL_allaxis} respectively.  
We see that overall, even if there are discrepancies due to the choice of the line of sight, the main trends remain the same as before.

\section{Local-Lagrangian approximation}

\begin{figure*}[htbp!]
\centering
\includegraphics[width=1.0\hsize]{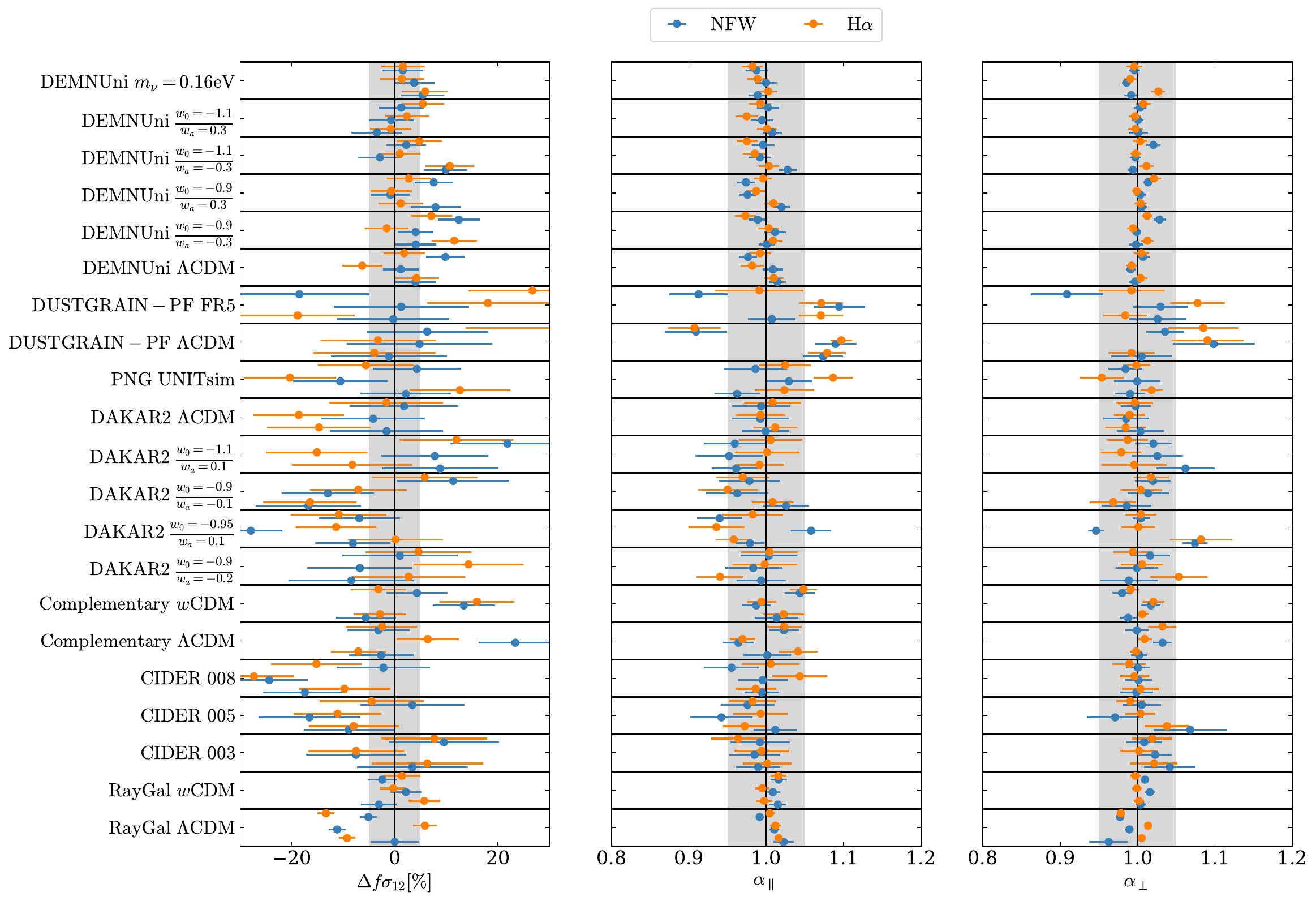}
\caption{Same as Fig.~\ref{fig:results_template_same_LL_allaxis} but without assuming the local-Lagrangian approximation.}
\label{fig:results_template_same_nonLL_allaxis}
\end{figure*}

\begin{figure*}[htbp!]
\centering
\includegraphics[width=1.0\hsize]{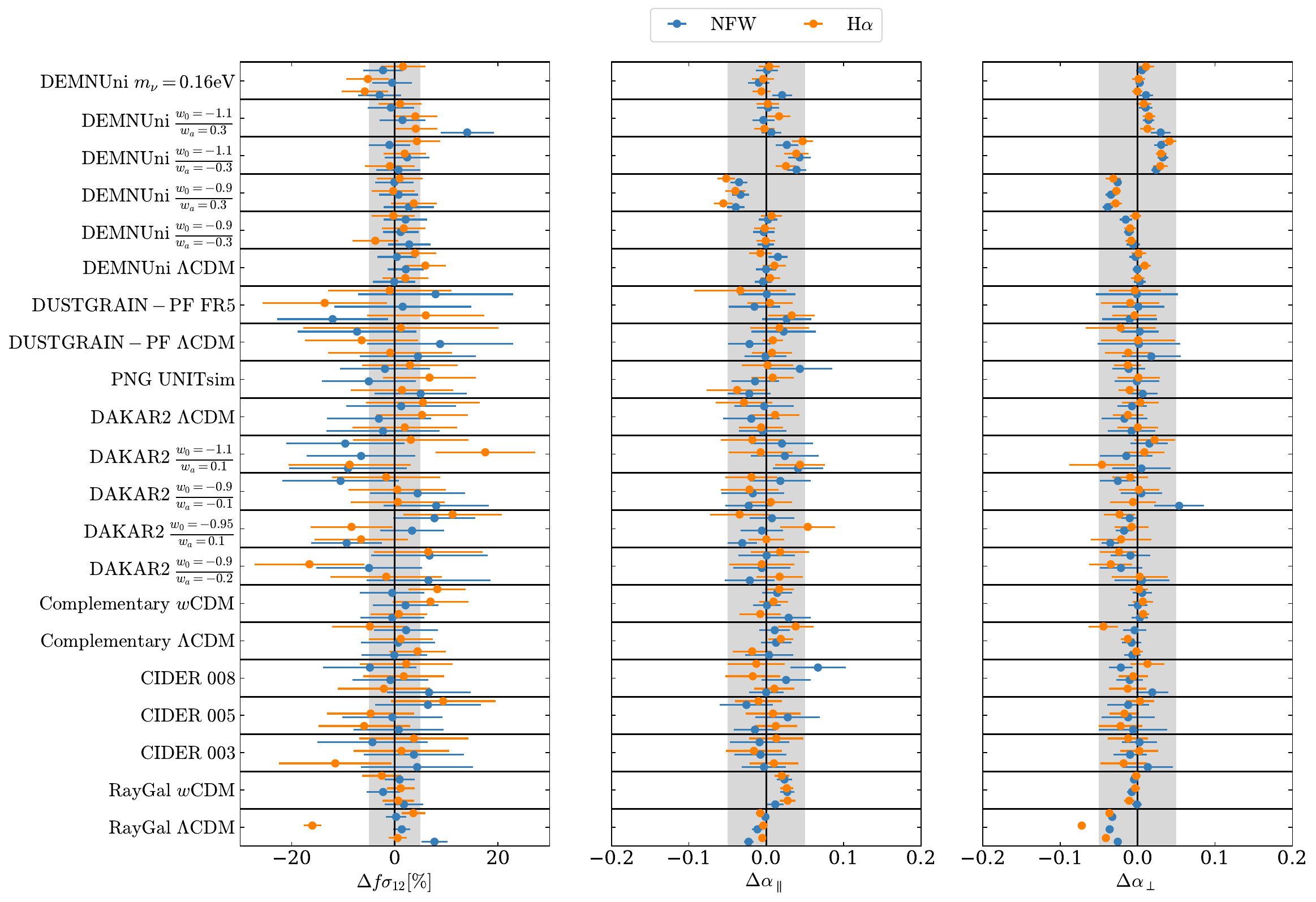}
\caption{Same as Fig.~\ref{fig:results_template_fs2_LL_allaxis} but without assuming the local-Lagrangian approximation.}
\label{fig:results_template_fs2_nonLL_allaxis}
\end{figure*}

In this final Appendix, we assess the impact of the local-Lagrangian approximation employed in the main analysis. In this approximation, we set the non-local bias terms to the relations shown in Eq.~\eqref{eq:LL_approximation}. We chose to use this approximation to minimise the number of free parameters involved in this analysis, and because it was shown in \citet{karcher20262PCF} that it did not have a strong impact on the parameter inference. We show the results for all lines of sight in Figs.~\ref{fig:results_template_same_nonLL_allaxis} and \ref{fig:results_template_fs2_nonLL_allaxis}.
We see that the results for AP parameters are very similar, although with more scatter. The main difference appears in the estimation of $f\sigma_{12}$. We do not clearly recover the expectation value due to an even larger scatter. Interestingly, there is now a clear difference between the NFW and H$\alpha$ profiles, which was not apparent in Figs.~\ref{fig:results_template_same_LL_allaxis} and \ref{fig:results_template_fs2_LL_allaxis}. This means that there is a non-trivial relation between halo mass profile and stochastic noise due to the catalogue peculiar realisation, that is captured in the non-local biases. This study is beyond the scope of the present paper and is left for future work. 

\end{appendix}

\end{document}